\documentclass[aps,pre,reprint]{revtex4-1}
\usepackage{amssymb}
\usepackage{graphicx}
\usepackage{dcolumn}
\usepackage{amsmath}
\usepackage{bm}
\usepackage{hyperref}
\usepackage{latexsym}
\usepackage{verbatim}
\usepackage{color}
\usepackage{subfigure}
\usepackage{bm}
\usepackage{xspace}
\usepackage{xr}

\newcommand{\rev}[1]{\textcolor{black}{#1}}

\newcommand{\mr}{\mathbf{r}}

\def\eg{\textit{e.g.}\xspace}
\newcommand{\etal}{et al.\xspace}

\def\fa{{f^\text{a}}}

\def\ktp{\kappa_\text{eff}}
\def\phd{+\frac{1}{2}}
\def\mhd{-\frac{1}{2}}

\def\Ndefect{n_\text{defect}}
\def\Dbend{D_\text{bend}}
\def\Dsplay{D_\text{splay}}
\def\dbend{\mathbf{d}_\text{bend}}
\def\dsplay{d_\text{splay}}

\begin{document}


\title{The interplay between activity and filament flexibility determines the emergent properties of active nematics}

\author{Abhijeet Joshi}
\email{email:aajoshi@brandeis.edu}
\affiliation{Martin Fisher school of Physics, Brandeis University, Waltham, MA 02453, USA}
\author{Elias Putzig}
\affiliation{Martin Fisher school of Physics, Brandeis University, Waltham, MA 02453, USA}
\author{Aparna Baskaran}
\email{email:aparna@brandeis.edu}
\affiliation{Martin Fisher school of Physics, Brandeis University, Waltham, MA 02453, USA}
\author{Michael F. Hagan}
\email{email:hagan@brandeis.edu}
\affiliation{Martin Fisher school of Physics, Brandeis University, Waltham, MA 02453, USA}

\date{\today}

\begin{abstract}
Active nematics are microscopically driven liquid crystals that exhibit dynamical steady states characterized by the creation and annihilation of topological defects. Motivated by experimental realizations of such systems made of biopolymer filaments and motor proteins, we describe a large-scale simulation study of a particle-based computational model that explicitly incorporates the semiflexibility of the biopolymers. We find that energy injected into the system at the particle scale preferentially excites bend deformations, renormalizing the filament bend modulus to smaller values. The emergent characteristics of the active nematic depend on activity and flexibility only through this activity-renormalized bend modulus, demonstrating that material parameters such as the Frank `constants' must explicitly depend on activity in a continuum hydrodynamic description of an active nematic. Further, we present a systematic way to estimate these material parameters from observations of deformation fields and defect shapes in experimental or simulation data.
\end{abstract}

\maketitle


Active nematics are a class of non-equilibrium liquid crystalline materials in which the constituent particles transduce energy into stress and motion, driving  orientationally inhomogeneous dynamical steady states characterized by generation, annihilation, and streaming of topological defects \cite{Sanchez2012, DeCamp2015, peng2016command, Zhou28012014, Narayan2007, narayan2006nonequilibrium}.
Being intrinsically far-from-equilibrium, active nematics enable developing materials with capabilities that would be thermodynamically forbidden in an equilibrium system. However, such applications require understanding how an active nematic's macroscale material properties depend on the microscale features of its constituent nematogens.

\begin{figure}[!h]
\includegraphics[width=0.35\textwidth]{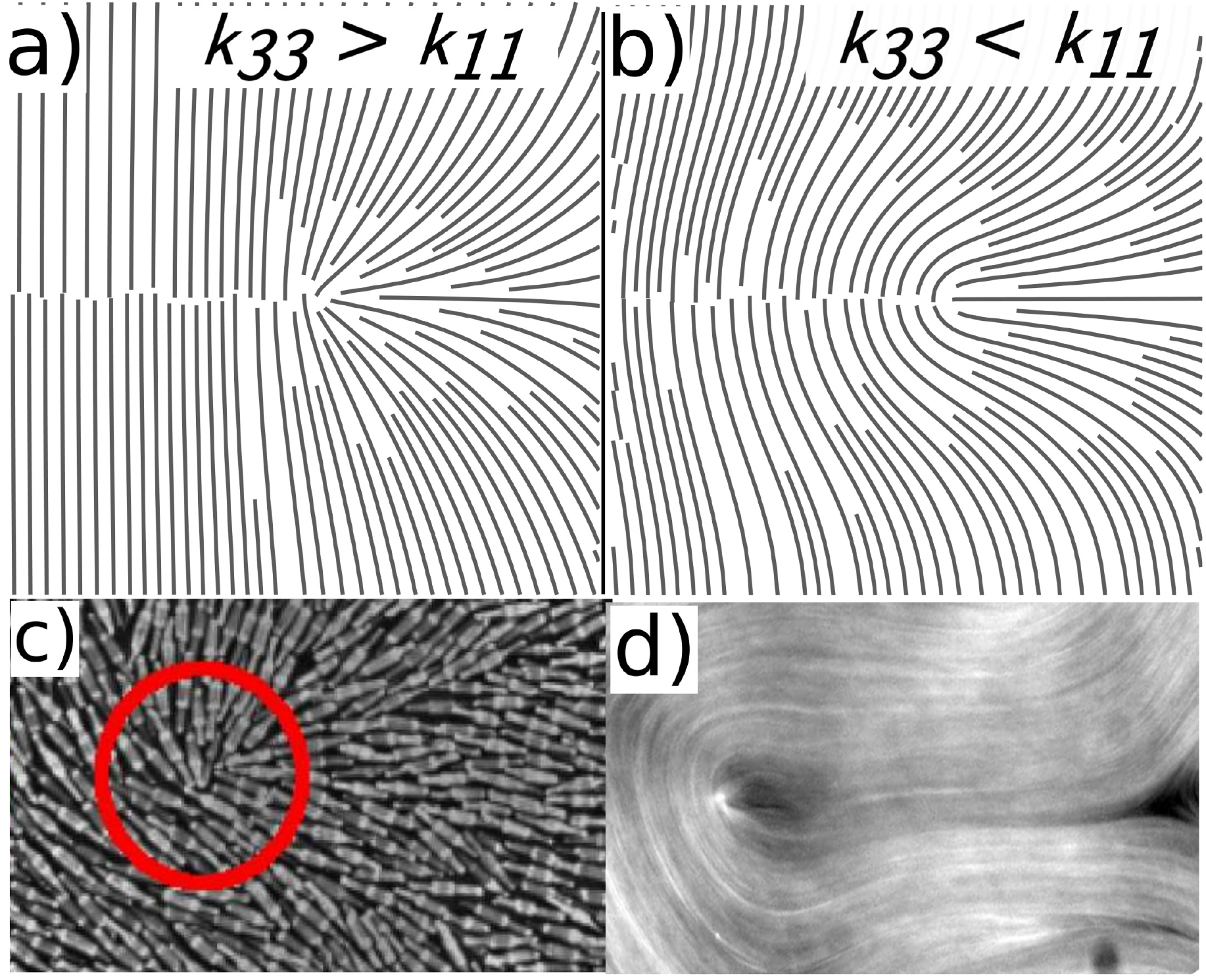}
\caption{\label{fig:dshapes_illustration}  {\bf (a,b)} The equilibrium director field
configuration near a $\phd$ defect for {\bf (a)} bend modulus larger
than splay modulus $k_{33}/k_{11}=3.4$,
leading to pointed defects and {\bf (b)} splay modulus larger than
bend $k_{33}/k_{11}=0.3$ leading to rounded defects. In each case
the director field was calculated by numerically minimizing the Frank
free energy around a separated $\pm \frac{1}{2}$ defect pair. {\bf (c,d)} Typical defect
morphologies observed in {\bf (c)} a vibrated granular nematic \cite{Narayan2007} and {\bf (d)} a microtubule-based active nematic \cite{Sanchez2012,Metcalf2017}.
}\label{Fig1}
\end{figure}

The challenge in this task can be illustrated by considering the shape of a defect. Defects have been the subject of intense research in equilibrium nematics, since they must be eliminated for display technologies, and controlled for applications such as directed assembly and biosensing \cite{Broer2010,Lin2011,Kleman1989}. Theory and experiments \cite{Kleman1989,Zhou2017,Zhang2018} have shown that defect morphologies depend on the relative values of the bend and splay elastic moduli, $k_{33}$ and $k_{11}$ (Fig.~\ref{Fig1}a),
\rev{defined in the Frank free energy density for a 2D nematic as  $f_\text{F}=k_{11} \left(\nabla \cdot \hat{n}\right)^2 + k_{33}\left(\hat{n} \times (\nabla \times \hat{n})\right)^2$ with $\hat{n}$ the director field \cite{Frank1958,Oseen1933}}.
Different experimental realizations of active nematics also exhibit variations in defect morphologies (\eg Fig.~\ref{Fig1}c and Fig.~\ref{Fig1}d). However, relating these defect morphologies
to material constants is much less straightforward than in equilibrium systems. Since moduli cannot be rigorously defined
in a non-equilibrium system such as an active nematic, there is no systematic way to measure them.

Computational and theoretical modeling could overcome these limitations.
Symmetry-based hydrodynamic theories \cite{Simha2002,Marchetti2013,Ramaswamy2010} have led to numerous insights about active nematics, including describing defect dynamics (\eg \cite{Giomi2013,Giomi2014,Doostmohammadi2017,Oza2016,Cortese2018}), induced flows in the suspending fluid (\eg \cite{Thampi2014,giomi2015geometry}), and how confinement in planar \cite{shendruk2017dancing,Norton2018,Gao2017} and curved geometries \cite{Keber2014,Zhang2016,Ellis2017,Alaimo2017} controls defect profileration and dynamics. However, \rev{hydrodynamic theories cannot predict how material constants or emergent behaviors depend on the microscale properties of individual nematogens}.
Further, while simulations of active nematics composed of rigid rods \cite{DeCamp2015,Gao2015,Shi2013} have elucidated emergent morphologies, \rev{these models do not account for internal degrees of freedom available to flexible nematogens \cite{Sanchez2012,sumino2012large,Guillamat2016}.}

\begin{figure}[!h]
\includegraphics[width=0.5\textwidth]{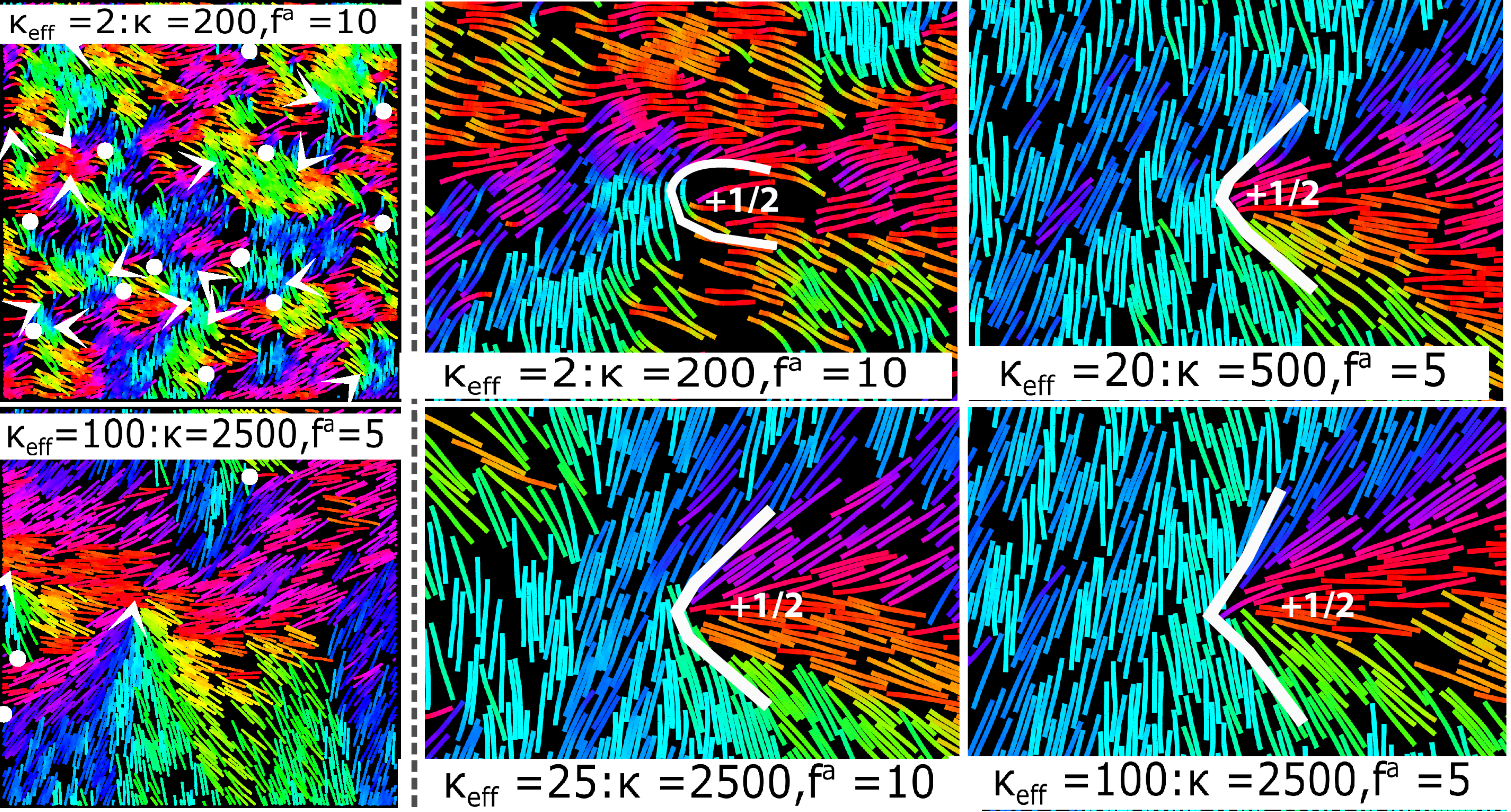}
\caption{\label{fig:snaps_illust}
\rev{A visual summary of how active nematic emergent properties depend on the activity parameter $\fa$ and the filament modulus $\kappa$.  The left panel in each row shows 1/16$^{th}$ of the simulation box for indicated parameter values, with white arrows indicating positions and orientations of $\phd$ defects and white dots indicating positions of $\mhd$ defects. Filament beads are colored according to the orientations of the local tangent vector. The right 4 panels are each zoomed in on a $\phd$ defect, with indicated parameter values. The white lines are drawn by eye to highlight defect shapes. Animations of corresponding simulation trajectories are in \cite{Supplement}.}
}\label{Fig2}
\end{figure}

In this work, we perform large-scale simulations on a model active nematic composed of semiflexible filaments, and determine how its emergent morphology depends on filament stiffness and activity. Fig.~\ref{Fig2} shows representative simulation outcomes. We find that characteristics of the active nematic texture --- the defect density, magnitudes and characteristic scales of bend and splay deformations, the shape of motile $\phd$ defects, \rev{and number fluctuations}  --- can all be described as  functions of an `activity-renormalized' nematogen rigidity, $\ktp\cong\kappa/(\fa)^2$, where $\kappa$ is the intrinsic bending modulus of the nematogen and $\fa$ is a measure of its activity. \rev{We show that one can define effective bend and splay moduli analogous to their equilibrium counterparts, but that the bend modulus is proportional to $\ktp$.    Thus, activity leads to an apparent softening of the nematic.
 This observation demonstrates that activity preferentially dissipates into particular modes within an active material, depending on the internal degrees of freedom of its constituent units. Moreover, these results suggest revisiting assumptions underlying hydrodynamic theories of active matter; for instance  material parameters should explicitly depend on activity.}

\rev{We also present a method of parameterizing defect shapes, which enables   estimating the relative magnitudes of renormalized bend and splay constants from experimental observations of defects, and thus allows direct experimental testing of our prediction. }



\textit{Model.}
We seek a minimal computational model that captures the key material properties and symmetries of active nematics, namely extensile activity, nematic inter-particle alignment and semiflexibility. We model active filaments as semiflexible bead-spring polymers, each containing $M$ beads of diameter $\sigma$, with neighboring beads connected by finitely extensible (FENE) bonds \cite{fene_1990}, and flexibility controlled by a harmonic angle potential between successive bonds that sets the filament bending modulus $\kappa$. \rev{Activity is implemented as a force of magnitude $\fa$ acting on each filament bead along the local tangent to the filament. To make activity nematic, the active force on every bead within a given filament reverses direction after a time interval chosen from a Poisson distribution with mean $\tau_1$.} Each bead is further subject to an isotropic stochastic white noise. Further model details are in Section S2 of \cite{Supplement}. The Langevin equations governing the system dynamics were integrated using a modified version of LAMMPS \cite{Plimpton1995}. \rev{The model combines features of recent models for polar self-propelled polymers ~\cite{Prathyusha2018} and the reversing rods approach to rigid active nematic rods~\cite{Shi2013}.}

\begin{figure*}
\includegraphics[width=0.99\textwidth]{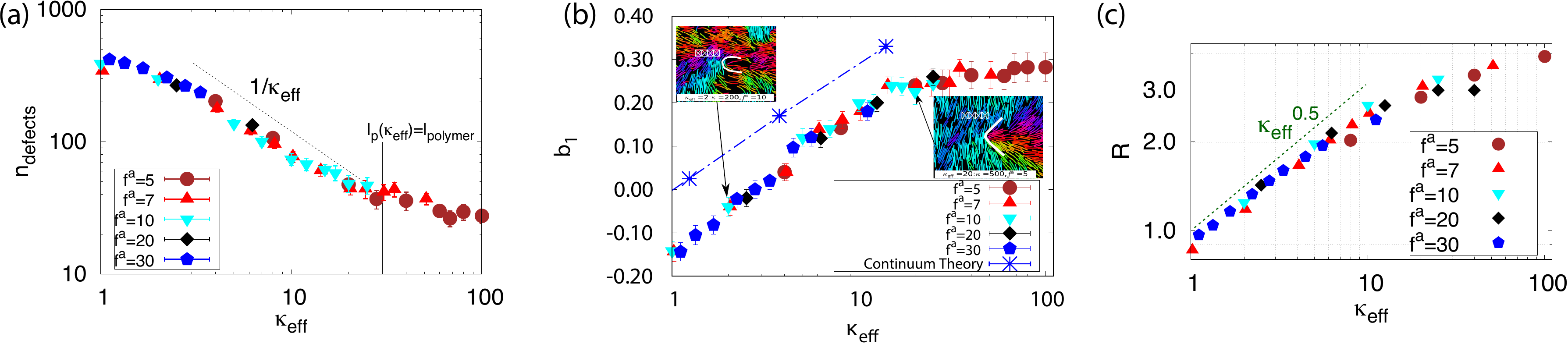}
\caption{\label{Fig3} \rev{The characteristics of an active nematic  are determined by the renormalized bending rigidity, $\ktp=\kappa/({\fa})^2$. \textbf{(a)} The average number of defects $\Ndefect$ observed in steady states as a function of $\ktp$, for activity parameter values $\fa \in [5,30]$ and $\kappa \in [100,3000]$.
\textbf{(b)} Mode of defect shape parameter $b_1$  for $\phd$ defects  as a function of $\ktp$ for $\fa \in [7,30]$ and $\kappa \in [100,10000]$, with   $b_1(r)$  values evaluated at
a radial distance $r=12.6\sigma$ from the defect core. The dashed
line and blue asterisk symbols show values of $b_1$ calculated for
isolated defects from equilibrium continuum elastic theory, with the ratio of
bend and splay moduli $k_{33}/k_{11}$ for each $\ktp$ set according to
the measured ratio $R$ of bend and splay deformations in (c). \textbf{(c)} The ratio of splay and bend deformations, ${\emph{R}}=\langle \int d^2\mr \Dsplay(\mr)\rangle / \langle \int d^2\mr \Dbend(\mr)\rangle$ as a function of $\ktp$ for the same parameters as in (b).  The simulation box size is  ($840\times840 \sigma^2$) for all cases. Explicit parameter values for each symbol are shown in Fig. S2 \cite{Supplement}. }
}
\end{figure*}

\rev{Since we are motivated by systems sufficiently large that boundary effects are negligible (\eg Ref. \cite{DeCamp2015})}, we performed simulations in a $A=840\times840\sigma^2$ periodic simulation box, with $N=19404$ filaments, each with $M=20$ beads, so that the packing fraction $\phi=MN\sigma^2/A \approx 0.55$. Fixing all other model parameters, we varied the filament stiffness, \rev{
$\kappa \in [10^2,10^4] k_\text{B}T_\text{ref}$ and the magnitude of the active force, $\fa \in [5,30] k_\text{B}T_\text{ref}/\sigma$, where the units are set by the thermal energy at a reference state $k_\text{B}T_\text{ref}$}, and the filament diameter $\sigma$. We initialized the system by first placing the filaments in completely extended configurations,
on a rectangular lattice with filaments oriented along one of the lattice vectors.
We then relaxed the initial configuration for $10^4\tau$ \rev{(by which point the defect density saturated for all parameter sets)}, where $\tau=\sqrt{m\sigma^2/k_\text{B}T_\text{ref}}$, before performing production runs of $\sim4\times10^4\tau$.
In the limit of high stiffness or low activity, the unphysical crystalline initial condition did not relax on computationally accessible timescales. In these cases, we used an alternative initial configuration, with filaments arranged into four rectangular
lattices, each placed in one quadrant of the simulation box such that adjacent (non-diagonal) lattices were orthogonal.

We analyzed simulation trajectories by calculating the local nematic tensor $Q_{\alpha\beta}$ and the density field $\rho$ on a $1000 \times 1000$ grid.
We then evaluated the local nematic order parameter $S$ and director $\hat{n}$ as the largest eigenvalue of $Q$ and corresponding eigenvector \rev{\cite{Eppenga1984}}. We  determined defect locations and orientations following standard approaches; further details are in Section S9 of \cite{Supplement}. \rev{We could not measure defect densities for systems with $\fa\ge20$ and $\kappa \gtrsim 5000$ because these systems exhibit localized density fluctuations which cause the detection algorithm to break down.}

\textit{Results.}
To understand how the interplay between filament flexibility and activity determines the texture of an active nematic, let us first consider the basic physics governing the emergent phenomenology in our system. Deforming a nematogen causes an elastic stress that scales $\thicksim \kappa$, the filament bending modulus. This elastic stress must be balanced by stresses arising from interparticle collisions. These collisional stresses  have two contributions. First there are the passive collisional stresses, \rev{which are $\mathcal{O}(1)$ in our units (as we have chosen the thermal energy and the particle size $\sigma$ as the energy and length scales)}. Then there are active collisional stresses that arise due to the self-replenishing active force of magnitude $\fa$ acting on each monomer. The active collisional stress will scale as $\nu_\text{coll} \Delta p_\text{coll}$, i.e., the collision frequency times the momentum transfer at each collision. Both the collision frequency  and momentum transfer will scale with the self-propulsion velocity \cite{Baskaran2008a}, which in our model is set by $\fa$. Thus, the collisional stresses will scale  $\thicksim(1+(\fa)^2)$.  \rev{So, the system properties  will be controlled by a parameter which measures the ratio of elastic to collisional stresses, $\ktp\equiv\kappa/(1+(\fa)^2)$.  Because we focus on the moderate to large activity regime, the passive stresses are negligible and we set $\ktp \approx \kappa/(\fa)^2$.}

We now test this argument against our simulation results, using several metrics for deformations in the nematic order:  defect density, defect shape, and the distribution of energy in splay and bend deformations. As shown in Fig.~\ref{Fig3}, all of these quantities depend on bending rigidity and activity only through the combination $\ktp$.


\textit{Defect Density:} We first test this argument in the context of a
 well-studied  property of an
active nematic, the steady-state defect density \cite{Giomi2014,Thampi2014,Putzig2015,Norton2018}. Fig.~\ref{Fig3}a plots
the defect density for a range of microscopic flexibility and activity values. \rev{The data collapses into a single curve, substantiating the arguments above.  Data collapse breaks down for $\fa<5$ (Fig. S1 \cite{Supplement}) when the system loses nematic order (see below).}
\rev{The defect density roughly scales as $1/\ktp$ under parameter values for which the renormalized persistence length lies between the monomer scale and filament contour length,  $1.5 \lesssim \ktp \lesssim \ktp^\text{max}$, with $\ktp^\text{max}\approx30$. While longer filaments would be required to rigorously test the scaling of defect density with $\ktp$, we found the same dependence over a wider parameter range by estimating the persistence length from filament tangent fluctuations  (Fig. S3, \cite{Supplement}), and we observe the same scaling for an additional data set (Fig. S1 \cite{Supplement}).} Hydrodynamic theories such as \cite{Giomi2014,Thampi2014,Putzig2015,Norton2018} predict
that the defect density scales as $\alpha/K$, where $\alpha$ is the measure of
active stress in the hydrodynamic theories and $K$ is a Frank elastic constant.
The scaling collapse observed in our simulations indicates that these
hydrodynamic parameters are directly related to their microscopic
analogs, namely the filament elastic constant $\kappa$ and the
collisional active stress $\sim(1+(\fa)^2)$.

\textit{Defect Shape:} Let us now consider a striking feature of the texture
of an active nematic, the shape of a $\phd$ defect. A quantitative
descriptor of defect shape can be developed as follows. We choose a
coordinate system with its origin at the center of the defect core, with the $x-$axis
pointing along the defect orientation (see Fig. S13 in \cite{Supplement}). We work in
polar coordinates, with $r$ the radial distance from the defect core and $\phi$ as the
azimuthal angle with respect to the $x-$axis. We define $\theta$ as the angle of the director field with respect to the $x-$axis. At any position, we can then express $\theta(r,\phi)$ as a Fourier expansion in $\phi \in \{-\pi:\pi\}$.
However, in practice truncating the expansion after the first sin term approximates the shape of a $\phd$ defect well, so the descriptor simplifies to
$\theta_{+\frac{1}{2}}(r, \ensuremath{\phi})\approx\ensuremath{\frac{1}{2}\phi}+\ensuremath{b_{1} (r)\sin}(\ensuremath{\phi})$,
\rev{and  the result is insensitive to $r$ provided the measurement is performed outside of the defect core (see Fig. S14 \cite{Supplement})}.
Thus, the parameter $b_1$ uniquely describes the defect shape; in particular, larger values of $b_1$ correspond to pointier defects. Note that Zhang \etal \rev{\cite{Zhang2018}} and Cortese \etal \cite{Cortese2018} proposed similar approaches to characterizing defect shapes while we were preparing this manuscript. An advantage of our approach is that the shape can be described by a single number, $b_1$.

Figure~\ref{Fig3}b shows the mode of $b_1$ calculated from
defects within our simulations depends only on   $\ktp$, thus demonstrating that defect shape is controlled by
the renormalized bending rigidity. Moreover, $b_1$ increases
logarithmically with $\ktp$; i.e., defects become more pointed as
the bending rigidity increases, consistent with the previous
observations of highly pointed defects in
rigid rod simulations \cite{Shi2014,DeCamp2015,Gao2015}.

\textit{Splay and Bend deformations:}  Next, to characterize the overall texture
of the active nematic, we consider the deformations in its
order field. In a 2D nematic, any arbitrary deformation of the director
field can be decomposed into a bend
deformation $\dbend=\left(\hat{n}(\mr) \times (\nabla \times \hat{n}(\mr))\right)$ and a
splay deformation $\dsplay=\left(\nabla\cdot\hat{n}(\mr)\right)$.  We define associated strain energy densities $\Dbend=\rho S^2 |\dbend|^2$ and $\Dsplay=\rho S^2 \dsplay^2$, which measure how the system's elastic energy distributes into the two linearly independent deformation modes. The prefactors of density $\rho$ and order $S$ allow the definition to be valid in regions such as defect cores and voids that occur in the particle simulations.

To  assess $\ktp$ as the analog of an equilibrium modulus,
let us consider the ratio of total strain energy in splay deformations to those in bend, ${\emph{R}}=\langle \int d^2\mr \Dsplay(\mr)\rangle / \langle \int d^2\mr \Dbend(\mr)\rangle$. In equilibrium, equipartition requires that this ratio satisfy $R_\text{eq}=k_{33}/k_{11}$. In our simulations we observe  data collapse for all parameter sets, with the ratio roughly scaling as ${\emph{R}}\thicksim \ktp^{1/2}$  (Fig.~\ref{Fig3}c). This result supports the existence of effective elastic moduli in an active nematic, but again shows that their values are activity-dependent. Note that this differs from the scaling of an equilibrium nematic, where the moduli scale as $k_{33}\thicksim \kappa$ and $k_{11} \thicksim \kappa^{1/3}$ \cite{Selinger1991,Dijkstra1995}, giving ${\emph{R}}_\text{eq}\thicksim \kappa^{2/3}$. We confirmed this scaling by calculating the equilibrium moduli for our system (Fig. S7 \cite{Supplement}) using the free energy perturbation technique of \cite{Joshi2014}. 


\begin{figure}[!t]
\includegraphics[width=0.99\columnwidth]{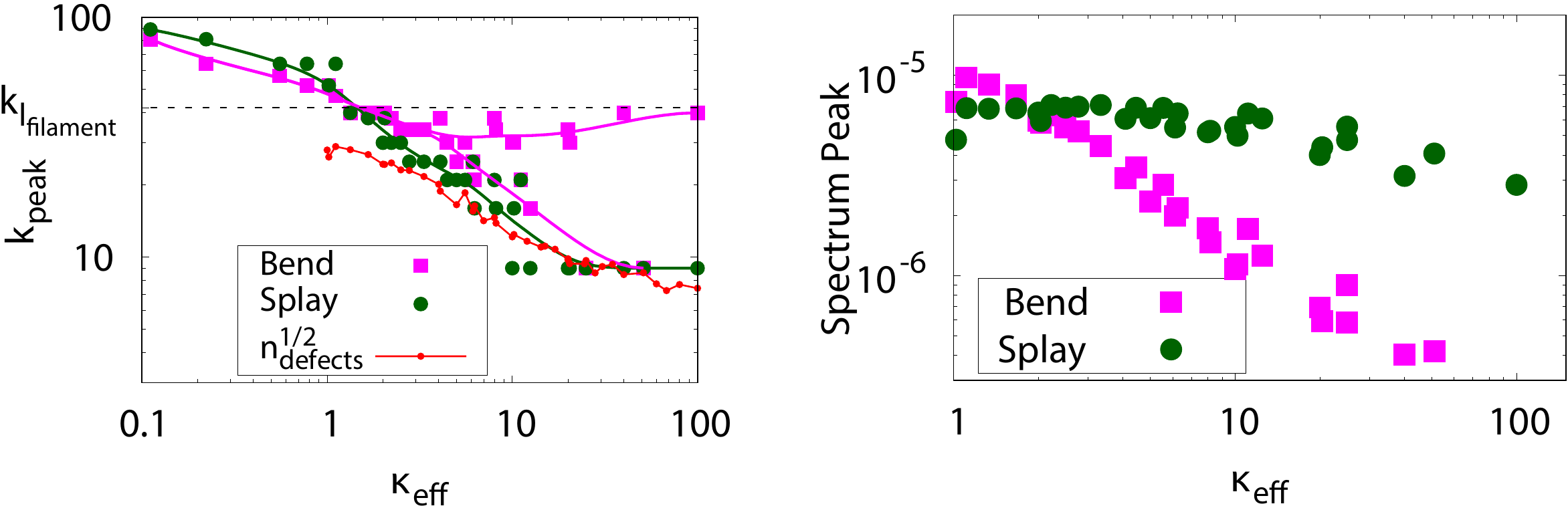}
\caption{Analysis of the power spectra of splay and bend deformation energy densities: {\bf (a)} The wavenumbers corresponding to peaks $k_\text{peak}$ in bend (\textcolor[rgb]{1.00,0.00,1.00}{$\blacksquare$} symbols) and
splay (\textcolor[rgb]{0.00,0.50,0.00}{\large{$\bullet$}} symbols) power spectra,
and  the inverse defect spacing, $\sqrt{\langle \Ndefect \rangle}$ (\textcolor{red}{\tiny{$\bullet$}} symbols) as
functions of $\ktp$.  The predominant scale of splay deformations transitions from the individual
filament scale ($k_\text{filament}$, dashed line) at low $\ktp$ to the defect spacing
scale at high $\ktp$. Bend spectra exhibit two peaks, corresponding to filament and defect-spacing scales, for high $\ktp$. {\bf (b)} The maximum magnitude of the power spectra, showing that splay is insensitive to $\ktp$ while bend
deformations decrease in magnitude until the effective rigidity exceeds the filament length.
\label{Fig4}
}
\end{figure}

Analysis of simulation trajectories suggests that the difference in scaling arises because the active systems have much higher nematic order than the equilibrium systems (when compared at the same values of renormalized bending rigidity).  In an equilibrium system, increasing the filament flexibility decreases order, resulting in a transition
from the nematic to isotropic phase as $\kappa$ decreases below $5$ at the density used in our simulations \footnote{Note that this 2D system with soft-core interactions should exhibit quasi-long-range order in the nematic phase
~\cite{Straley1971,FrenkelEppenga1985,Kosterlitz1973},
and thus $S$  decays with system size. Since we are interested in the degree of order on the scale of
typical bend and splay fluctuations, we report $S$ values measured within regions of size $10\times10 \sigma^2$ (see \cite{Supplement}).}.
In contrast, the active systems remain deep within the nematic phase ($S \gtrsim 0.8$) for all parameter
sets (Fig. S7c in \cite{Supplement}).
 We empirically found that normalizing the bend-splay ratio by the degree of order, using the form ${\emph{R}}/S^2$, roughly collapses the equilibrium and active results onto a single curve (Fig. S7b \cite{Supplement}). Although we cannot rationalize the form of the normalization, this result implies that the differences in scaling between active and equilibrium cases arise from the increased nematic order in the active simulations.


The power spectra associated with these strain energies (Fig. S6 in \cite{Supplement})  exhibit two features: a) Fig. \ref{Fig4}a shows that for the smallest simulated effective rigidity, $\ktp=2$, the peak of the power spectrum $k_\text{peak}$ is on the filament scale, indicating that most deformation energy arises due to bending of individual filaments. But for all higher values of $\ktp$, a peak arises at the defect spacing scale, indicating that the defect density  controls the texture in this active nematic. 
b) Fig. \ref{Fig4}b shows the magnitude of the power spectrum at the characteristic scale $k_\text{peak}$. The splay energy density is nearly independent of the effective filament stiffness, while the amount of bend decreases linearly with $\ktp$ (for $\ktp\lesssim\ktp^\text{max}$). This observation further supports  $\ktp$ as an effective bend modulus, and demonstrates that energy injected into the system at the microscale due to activity preferentially dissipates into bend modes.

Returning to the defect shape analysis, the dashed line in Fig.~\ref{Fig3}b shows the defect shape parameter values $b_1$ calculated by minimizing the continuum Frank free energy for an equilibrium system containing a separated pair of $\phd$ and $\mhd$ defects. Here we have fixed the continuum modulus $k_{11}$ and varied $k_{33}$ so that $\emph{R}_\text{eq}=\frac{k_{33}}{k_{11}}$ matches the value obtained from our microscopic simulations at the corresponding value of effective stiffness:  $\emph{R}_\text{eq}(k_{33})=R(\ktp)$. Although the equilibrium $b_1$ values are shifted above the simulation results, consistent with the discrepancy in $\emph{R}$ between the active and passive systems (Fig.~\ref{Fig3}C and Fig. S7 \cite{Supplement}), we observe the same scaling with bending rigidity in the continuum and simulation results. Thus, the defect shape measurement provides a straightforward means to estimate effective moduli in simulations or experiments on active nematics.

\rev{Finally, previous work has shown that active nematics are susceptible to phase separation \cite{Mishra2006,Ngo2014,Putzig2014,Shi2013} and giant number fluctuations \cite{Ramaswamy2003,Mishra2006,Chate2006,Narayan2007,Zhang2010}. We find that the renormalized modulus governs the latter, with deviations from equilibrium scaling roughly proportional to $\ktp$. However, flexibility suppresses phase separation except at molecular scales (Fig. S11 \cite{Supplement}).} 

\textit{Conclusions.} In active materials, energy injected at the particle scale undergoes an inverse cascade, manifesting in collective spatiotemporal dynamics at larger scales. The nature of this emergent behavior depends crucially on which modes the active energy dissipates into. Our simulations show that activity within a dry active nematic preferentially dissipates into bend modes, thus  softening  the apparent bend modulus. This renormalized bend modulus dictates all collective behaviors that we investigated ---  the relative amounts of bend and splay, defect density, defect shape, \rev{and giant number fluctuations}.
The dependence of multiple characteristics on this single parameter suggests that bend and splay moduli can be meaningfully assessed in an active nematic, despite the fact that there is no rigorous definition for a modulus in a non-equilibrium active material. Finally, our technique to parameterize defect shapes allows estimating the relative magnitudes of these effective moduli in any experimental or computational system that exhibits defects.

\begin{acknowledgments}
This work was supported by the Brandeis Center for Bioinspired Soft Materials, an NSF MRSEC,  DMR-1420382 (AJ,EP,AB,MFH) and NSF DMR-1149266 (AB). Computational resources were provided by NSF XSEDE computing resources (MCB090163, Stampede) and the Brandeis HPCC which is partially supported by DMR-1420382.
\end{acknowledgments}


\begin{thebibliography}{23}%
\makeatletter
\providecommand \@ifxundefined [1]{%
 \@ifx{#1\undefined}
}%
\providecommand \@ifnum [1]{%
 \ifnum #1\expandafter \@firstoftwo
 \else \expandafter \@secondoftwo
 \fi
}%
\providecommand \@ifx [1]{%
 \ifx #1\expandafter \@firstoftwo
 \else \expandafter \@secondoftwo
 \fi
}%
\providecommand \natexlab [1]{#1}%
\providecommand \enquote  [1]{``#1''}%
\providecommand \bibnamefont  [1]{#1}%
\providecommand \bibfnamefont [1]{#1}%
\providecommand \citenamefont [1]{#1}%
\providecommand \href@noop [0]{\@secondoftwo}%
\providecommand \href [0]{\begingroup \@sanitize@url \@href}%
\providecommand \@href[1]{\@@startlink{#1}\@@href}%
\providecommand \@@href[1]{\endgroup#1\@@endlink}%
\providecommand \@sanitize@url [0]{\catcode `\\12\catcode `\$12\catcode
  `\&12\catcode `\#12\catcode `\^12\catcode `\_12\catcode `\%12\relax}%
\providecommand \@@startlink[1]{}%
\providecommand \@@endlink[0]{}%
\providecommand \url  [0]{\begingroup\@sanitize@url \@url }%
\providecommand \@url [1]{\endgroup\@href {#1}{\urlprefix }}%
\providecommand \urlprefix  [0]{URL }%
\providecommand \Eprint [0]{\href }%
\providecommand \doibase [0]{http://dx.doi.org/}%
\providecommand \selectlanguage [0]{\@gobble}%
\providecommand \bibinfo  [0]{\@secondoftwo}%
\providecommand \bibfield  [0]{\@secondoftwo}%
\providecommand \translation [1]{[#1]}%
\providecommand \BibitemOpen [0]{}%
\providecommand \bibitemStop [0]{}%
\providecommand \bibitemNoStop [0]{.\EOS\space}%
\providecommand \EOS [0]{\spacefactor3000\relax}%
\providecommand \BibitemShut  [1]{\csname bibitem#1\endcsname}%
\let\auto@bib@innerbib\@empty
\bibitem [{\citenamefont {Weeks}\ \emph {et~al.}(1971)\citenamefont {Weeks},
  \citenamefont {Chandler},\ and\ \citenamefont {Andersen}}]{Weeks1971}%
  \BibitemOpen
  \bibfield  {author} {\bibinfo {author} {\bibfnamefont {J.~D.}\ \bibnamefont
  {Weeks}}, \bibinfo {author} {\bibfnamefont {D.}~\bibnamefont {Chandler}}, \
  and\ \bibinfo {author} {\bibfnamefont {H.~C.}\ \bibnamefont {Andersen}},\
  }\href@noop {} {\bibfield  {journal} {\bibinfo  {journal} {J. Chem. Phys.}\
  }\textbf {\bibinfo {volume} {54}},\ \bibinfo {pages} {5237} (\bibinfo {year}
  {1971})}\BibitemShut {NoStop}%
\bibitem [{\citenamefont {Kremer}\ and\ \citenamefont
  {Grest}(1990)}]{fene_1990}%
  \BibitemOpen
  \bibfield  {author} {\bibinfo {author} {\bibfnamefont {K.}~\bibnamefont
  {Kremer}}\ and\ \bibinfo {author} {\bibfnamefont {G.~S.}\ \bibnamefont
  {Grest}},\ }\href {\doibase 10.1063/1.458541} {\bibfield  {journal} {\bibinfo
   {journal} {J. Chem. Phys.}\ }\textbf {\bibinfo {volume} {92}},\ \bibinfo
  {pages} {5057} (\bibinfo {year} {1990})},\ \Eprint
  {http://arxiv.org/abs/http://dx.doi.org/10.1063/1.458541}
  {http://dx.doi.org/10.1063/1.458541} \BibitemShut {NoStop}%
\bibitem [{\citenamefont {Plimpton}(1995)}]{Plimpton1995}%
  \BibitemOpen
  \bibfield  {author} {\bibinfo {author} {\bibfnamefont {S.}~\bibnamefont
  {Plimpton}},\ }\href {\doibase 10.1006/jcph.1995.1039} {\bibfield  {journal}
  {\bibinfo  {journal} {J. Comput. Phys.}\ }\textbf {\bibinfo {volume} {117}},\
  \bibinfo {pages} {1} (\bibinfo {year} {1995})}\BibitemShut {NoStop}%
\bibitem [{\citenamefont {Peruani}\ \emph {et~al.}(2006)\citenamefont
  {Peruani}, \citenamefont {Deutsch},\ and\ \citenamefont
  {B\"ar}}]{Peruani2006}%
  \BibitemOpen
  \bibfield  {author} {\bibinfo {author} {\bibfnamefont {F.}~\bibnamefont
  {Peruani}}, \bibinfo {author} {\bibfnamefont {A.}~\bibnamefont {Deutsch}}, \
  and\ \bibinfo {author} {\bibfnamefont {M.}~\bibnamefont {B\"ar}},\ }\href
  {\doibase 10.1103/PhysRevE.74.030904} {\bibfield  {journal} {\bibinfo
  {journal} {Phys. Rev. E}\ }\textbf {\bibinfo {volume} {74}},\ \bibinfo
  {pages} {030904} (\bibinfo {year} {2006})}\BibitemShut {NoStop}%
\bibitem [{\citenamefont {{McCandlish}}\ \emph {et~al.}(2012)\citenamefont
  {{McCandlish}}, \citenamefont {Baskaran},\ and\ \citenamefont
  {Hagan}}]{McCandlish2012}%
  \BibitemOpen
  \bibfield  {author} {\bibinfo {author} {\bibfnamefont {S.~R.}\ \bibnamefont
  {{McCandlish}}}, \bibinfo {author} {\bibfnamefont {A.}~\bibnamefont
  {Baskaran}}, \ and\ \bibinfo {author} {\bibfnamefont {M.~F.}\ \bibnamefont
  {Hagan}},\ }\href {\doibase 10.1039/c2sm06960a} {\bibfield  {journal}
  {\bibinfo  {journal} {Soft Matter}\ }\textbf {\bibinfo {volume} {8}},\
  \bibinfo {pages} {2527} (\bibinfo {year} {2012})}\BibitemShut {NoStop}%
\bibitem [{\citenamefont {Weitz}\ \emph {et~al.}(2015)\citenamefont {Weitz},
  \citenamefont {Deutsch},\ and\ \citenamefont {Peruani}}]{Weitz2015}%
  \BibitemOpen
  \bibfield  {author} {\bibinfo {author} {\bibfnamefont {S.}~\bibnamefont
  {Weitz}}, \bibinfo {author} {\bibfnamefont {A.}~\bibnamefont {Deutsch}}, \
  and\ \bibinfo {author} {\bibfnamefont {F.}~\bibnamefont {Peruani}},\ }\href
  {\doibase 10.1103/PhysRevE.92.012322} {\bibfield  {journal} {\bibinfo
  {journal} {Phys. Rev. E}\ }\textbf {\bibinfo {volume} {92}},\ \bibinfo
  {pages} {012322} (\bibinfo {year} {2015})}\BibitemShut {NoStop}%
\bibitem [{\citenamefont {Ginelli}\ \emph {et~al.}(2010)\citenamefont
  {Ginelli}, \citenamefont {Peruani}, \citenamefont {B\"ar},\ and\
  \citenamefont {Chat\'{e}}}]{Ginelli2010}%
  \BibitemOpen
  \bibfield  {author} {\bibinfo {author} {\bibfnamefont {F.}~\bibnamefont
  {Ginelli}}, \bibinfo {author} {\bibfnamefont {F.}~\bibnamefont {Peruani}},
  \bibinfo {author} {\bibfnamefont {M.}~\bibnamefont {B\"ar}}, \ and\ \bibinfo
  {author} {\bibfnamefont {H.}~\bibnamefont {Chat\'{e}}},\ }\href {\doibase
  10.1103/PhysRevLett.104.184502} {\bibfield  {journal} {\bibinfo  {journal}
  {Phys. Rev. Lett.}\ }\textbf {\bibinfo {volume} {104}},\ \bibinfo {pages}
  {184502} (\bibinfo {year} {2010})}\BibitemShut {NoStop}%
\bibitem [{\citenamefont {Peruani}\ \emph {et~al.}(2012)\citenamefont
  {Peruani}, \citenamefont {Starru{\ss}}, \citenamefont {Jakovljevic},
  \citenamefont {S{\o}gaard-Andersen}, \citenamefont {Deutsch},\ and\
  \citenamefont {B\"{a}r}}]{Peruani2012}%
  \BibitemOpen
  \bibfield  {author} {\bibinfo {author} {\bibfnamefont {F.}~\bibnamefont
  {Peruani}}, \bibinfo {author} {\bibfnamefont {J.}~\bibnamefont
  {Starru{\ss}}}, \bibinfo {author} {\bibfnamefont {V.}~\bibnamefont
  {Jakovljevic}}, \bibinfo {author} {\bibfnamefont {L.}~\bibnamefont
  {S{\o}gaard-Andersen}}, \bibinfo {author} {\bibfnamefont {A.}~\bibnamefont
  {Deutsch}}, \ and\ \bibinfo {author} {\bibfnamefont {M.}~\bibnamefont
  {B\"{a}r}},\ }\href {\doibase 10.1103/PhysRevLett.108.098102} {\bibfield
  {journal} {\bibinfo  {journal} {Phys. Rev. Lett.}\ }\textbf {\bibinfo
  {volume} {108}},\ \bibinfo {pages} {098102} (\bibinfo {year}
  {2012})}\BibitemShut {NoStop}%
\bibitem [{\citenamefont {Joshi}\ \emph {et~al.}(2014)\citenamefont {Joshi},
  \citenamefont {Whitmer}, \citenamefont {Guzm{\'{a}}n}, \citenamefont
  {Abbott},\ and\ \citenamefont {de~Pablo}}]{Joshi2014}%
  \BibitemOpen
  \bibfield  {author} {\bibinfo {author} {\bibfnamefont {A.~A.}\ \bibnamefont
  {Joshi}}, \bibinfo {author} {\bibfnamefont {J.~K.}\ \bibnamefont {Whitmer}},
  \bibinfo {author} {\bibfnamefont {O.}~\bibnamefont {Guzm{\'{a}}n}}, \bibinfo
  {author} {\bibfnamefont {N.~L.}\ \bibnamefont {Abbott}}, \ and\ \bibinfo
  {author} {\bibfnamefont {J.~J.}\ \bibnamefont {de~Pablo}},\ }\href {\doibase
  10.1039/C3SM51919H} {\bibfield  {journal} {\bibinfo  {journal} {Soft Matter}\
  }\textbf {\bibinfo {volume} {10}},\ \bibinfo {pages} {882} (\bibinfo {year}
  {2014})}\BibitemShut {NoStop}%
\bibitem [{\citenamefont {Doi}\ and\ \citenamefont {Edwards}(1988)}]{Doi1988}%
  \BibitemOpen
  \bibfield  {author} {\bibinfo {author} {\bibfnamefont {M.}~\bibnamefont
  {Doi}}\ and\ \bibinfo {author} {\bibfnamefont {S.}~\bibnamefont {Edwards}},\
  }\href {https://books.google.com/books?id=Wie5ngEACAAJ} {\emph {\bibinfo
  {title} {The Theory of Polymer Dynamics}}},\ International series of
  monographs on physics\ (\bibinfo  {publisher} {Clarendon Press},\ \bibinfo
  {year} {1988})\BibitemShut {NoStop}%
\bibitem [{\citenamefont {Wensink}\ \emph {et~al.}(2012)\citenamefont
  {Wensink}, \citenamefont {Dunkel}, \citenamefont {Heidenreich}, \citenamefont
  {Drescher}, \citenamefont {Goldstein}, \citenamefont {Lowen},\ and\
  \citenamefont {Yeomans}}]{Wensink2012}%
  \BibitemOpen
  \bibfield  {author} {\bibinfo {author} {\bibfnamefont {H.~H.}\ \bibnamefont
  {Wensink}}, \bibinfo {author} {\bibfnamefont {J.}~\bibnamefont {Dunkel}},
  \bibinfo {author} {\bibfnamefont {S.}~\bibnamefont {Heidenreich}}, \bibinfo
  {author} {\bibfnamefont {K.}~\bibnamefont {Drescher}}, \bibinfo {author}
  {\bibfnamefont {R.~E.}\ \bibnamefont {Goldstein}}, \bibinfo {author}
  {\bibfnamefont {H.}~\bibnamefont {Lowen}}, \ and\ \bibinfo {author}
  {\bibfnamefont {J.~M.}\ \bibnamefont {Yeomans}},\ }\href {\doibase
  10.1073/pnas.1202032109} {\bibfield  {journal} {\bibinfo  {journal} {Proc.
  Nat. Acad. Sci. U.S.A.}\ }\textbf {\bibinfo {volume} {109}},\ \bibinfo
  {pages} {14308} (\bibinfo {year} {2012})},\ \Eprint
  {http://arxiv.org/abs/1208.4239v1} {1208.4239v1} \BibitemShut {NoStop}%
\bibitem [{\citenamefont {Chatterjee}\ \emph {et~al.}(2016)\citenamefont
  {Chatterjee}, \citenamefont {Joshi},\ and\ \citenamefont
  {Perlekar}}]{Chatterjee2016}%
  \BibitemOpen
  \bibfield  {author} {\bibinfo {author} {\bibfnamefont {R.}~\bibnamefont
  {Chatterjee}}, \bibinfo {author} {\bibfnamefont {A.~A.}\ \bibnamefont
  {Joshi}}, \ and\ \bibinfo {author} {\bibfnamefont {P.}~\bibnamefont
  {Perlekar}},\ }\href {\doibase 10.1103/PhysRevE.94.022406} {\bibfield
  {journal} {\bibinfo  {journal} {Phys. Rev. E}\ }\textbf {\bibinfo {volume}
  {94}},\ \bibinfo {pages} {022406} (\bibinfo {year} {2016})},\ \Eprint
  {http://arxiv.org/abs/1608.01142} {arXiv:1608.01142} \BibitemShut {NoStop}%
\bibitem [{\citenamefont {Mishra}\ and\ \citenamefont
  {Ramaswamy}(2006)}]{Mishra2006}%
  \BibitemOpen
  \bibfield  {author} {\bibinfo {author} {\bibfnamefont {S.}~\bibnamefont
  {Mishra}}\ and\ \bibinfo {author} {\bibfnamefont {S.}~\bibnamefont
  {Ramaswamy}},\ }\href {\doibase 10.1103/PhysRevLett.97.090602} {\bibfield
  {journal} {\bibinfo  {journal} {Phys. Rev. Lett.}\ }\textbf {\bibinfo
  {volume} {97}},\ \bibinfo {pages} {090602} (\bibinfo {year}
  {2006})}\BibitemShut {NoStop}%
\bibitem [{\citenamefont {Chat\'{e}}\ \emph {et~al.}(2006)\citenamefont
  {Chat\'{e}}, \citenamefont {Ginelli},\ and\ \citenamefont
  {Montagne}}]{Chate2006}%
  \BibitemOpen
  \bibfield  {author} {\bibinfo {author} {\bibfnamefont {H.}~\bibnamefont
  {Chat\'{e}}}, \bibinfo {author} {\bibfnamefont {F.}~\bibnamefont {Ginelli}},
  \ and\ \bibinfo {author} {\bibfnamefont {R.}~\bibnamefont {Montagne}},\
  }\href {\doibase 10.1103/PhysRevLett.96.180602} {\bibfield  {journal}
  {\bibinfo  {journal} {Phys. Rev. Lett}\ }\textbf {\bibinfo {volume} {96}},\
  \bibinfo {pages} {180602} (\bibinfo {year} {2006})}\BibitemShut {NoStop}%
\bibitem [{\citenamefont {Ngo}\ \emph {et~al.}(2014)\citenamefont {Ngo},
  \citenamefont {Peshkov}, \citenamefont {Aranson}, \citenamefont {Bertin},
  \citenamefont {Ginelli},\ and\ \citenamefont {Chat{\'{e}}}}]{Ngo2014}%
  \BibitemOpen
  \bibfield  {author} {\bibinfo {author} {\bibfnamefont {S.}~\bibnamefont
  {Ngo}}, \bibinfo {author} {\bibfnamefont {A.}~\bibnamefont {Peshkov}},
  \bibinfo {author} {\bibfnamefont {I.~S.}\ \bibnamefont {Aranson}}, \bibinfo
  {author} {\bibfnamefont {E.}~\bibnamefont {Bertin}}, \bibinfo {author}
  {\bibfnamefont {F.}~\bibnamefont {Ginelli}}, \ and\ \bibinfo {author}
  {\bibfnamefont {H.}~\bibnamefont {Chat{\'{e}}}},\ }\href {\doibase
  10.1103/PhysRevLett.113.038302} {\bibfield  {journal} {\bibinfo  {journal}
  {Phys. Rev. Lett.}\ }\textbf {\bibinfo {volume} {113}},\ \bibinfo {pages}
  {038302} (\bibinfo {year} {2014})},\ \Eprint {http://arxiv.org/abs/1312.1076}
  {arXiv:1312.1076} \BibitemShut {NoStop}%
\bibitem [{\citenamefont {Putzig}\ and\ \citenamefont
  {Baskaran}(2014)}]{Putzig2014}%
  \BibitemOpen
  \bibfield  {author} {\bibinfo {author} {\bibfnamefont {E.}~\bibnamefont
  {Putzig}}\ and\ \bibinfo {author} {\bibfnamefont {A.}~\bibnamefont
  {Baskaran}},\ }\href {\doibase 10.1103/PhysRevE.90.042304} {\bibfield
  {journal} {\bibinfo  {journal} {Phys. Rev. E}\ }\textbf {\bibinfo {volume}
  {90}},\ \bibinfo {pages} {042304} (\bibinfo {year} {2014})},\ \Eprint
  {http://arxiv.org/abs/1057984} {1057984} \BibitemShut {NoStop}%
\bibitem [{\citenamefont {Shi}\ and\ \citenamefont {Ma}(2013)}]{Shi2013}%
  \BibitemOpen
  \bibfield  {author} {\bibinfo {author} {\bibfnamefont {X.-q.}\ \bibnamefont
  {Shi}}\ and\ \bibinfo {author} {\bibfnamefont {Y.-q.}\ \bibnamefont {Ma}},\
  }\href {\doibase 10.1038/ncomms4013} {\bibfield  {journal} {\bibinfo
  {journal} {Nat. Comm.}\ }\textbf {\bibinfo {volume} {4}},\ \bibinfo {pages}
  {3013} (\bibinfo {year} {2013})}\BibitemShut {NoStop}%
\bibitem [{\citenamefont {Ramaswamy}\ \emph {et~al.}(2003)\citenamefont
  {Ramaswamy}, \citenamefont {Simha},\ and\ \citenamefont
  {Toner}}]{Ramaswamy2003}%
  \BibitemOpen
  \bibfield  {author} {\bibinfo {author} {\bibfnamefont {S.}~\bibnamefont
  {Ramaswamy}}, \bibinfo {author} {\bibfnamefont {R.~A.}\ \bibnamefont
  {Simha}}, \ and\ \bibinfo {author} {\bibfnamefont {J.}~\bibnamefont
  {Toner}},\ }\href {\doibase 10.1209/epl/i2003-00346-7} {\bibfield  {journal}
  {\bibinfo  {journal} {Europhys. Lett.}\ }\textbf {\bibinfo {volume} {62}},\
  \bibinfo {pages} {196} (\bibinfo {year} {2003})}\BibitemShut {NoStop}%
\bibitem [{\citenamefont {Narayan}\ \emph {et~al.}(2007)\citenamefont
  {Narayan}, \citenamefont {Ramaswamy},\ and\ \citenamefont
  {Menon}}]{Narayan2007}%
  \BibitemOpen
  \bibfield  {author} {\bibinfo {author} {\bibfnamefont {V.}~\bibnamefont
  {Narayan}}, \bibinfo {author} {\bibfnamefont {S.}~\bibnamefont {Ramaswamy}},
  \ and\ \bibinfo {author} {\bibfnamefont {N.}~\bibnamefont {Menon}},\ }\href
  {\doibase 10.1126/science.1140414} {\bibfield  {journal} {\bibinfo  {journal}
  {Science}\ }\textbf {\bibinfo {volume} {317}},\ \bibinfo {pages} {105}
  (\bibinfo {year} {2007})}\BibitemShut {NoStop}%
\bibitem [{\citenamefont {Zhang}\ \emph {et~al.}(2010)\citenamefont {Zhang},
  \citenamefont {Be{\textquoteright}er}, \citenamefont {Florin},\ and\
  \citenamefont {Swinney}}]{Zhang2010}%
  \BibitemOpen
  \bibfield  {author} {\bibinfo {author} {\bibfnamefont {H.~P.}\ \bibnamefont
  {Zhang}}, \bibinfo {author} {\bibfnamefont {A.}~\bibnamefont
  {Be{\textquoteright}er}}, \bibinfo {author} {\bibfnamefont {E.-L.}\
  \bibnamefont {Florin}}, \ and\ \bibinfo {author} {\bibfnamefont {H.~L.}\
  \bibnamefont {Swinney}},\ }\href {\doibase 10.1073/pnas.1001651107}
  {\bibfield  {journal} {\bibinfo  {journal} {Proc. Natl. Acad. Sci. U. S. A.}\
  }\textbf {\bibinfo {volume} {107}},\ \bibinfo {pages} {13626} (\bibinfo
  {year} {2010})},\ \Eprint
  {http://arxiv.org/abs/http://www.pnas.org/content/107/31/13626.full.pdf}
  {http://www.pnas.org/content/107/31/13626.full.pdf} \BibitemShut {NoStop}%
\bibitem [{\citenamefont {DeCamp}\ \emph {et~al.}(2015)\citenamefont {DeCamp},
  \citenamefont {Redner}, \citenamefont {Baskaran}, \citenamefont {Hagan},\
  and\ \citenamefont {Dogic}}]{DeCamp2015}%
  \BibitemOpen
  \bibfield  {author} {\bibinfo {author} {\bibfnamefont {S.~J.}\ \bibnamefont
  {DeCamp}}, \bibinfo {author} {\bibfnamefont {G.~S.}\ \bibnamefont {Redner}},
  \bibinfo {author} {\bibfnamefont {A.}~\bibnamefont {Baskaran}}, \bibinfo
  {author} {\bibfnamefont {M.~F.}\ \bibnamefont {Hagan}}, \ and\ \bibinfo
  {author} {\bibfnamefont {Z.}~\bibnamefont {Dogic}},\ }\href@noop {}
  {\bibfield  {journal} {\bibinfo  {journal} {Nat. Mater.}\ }\textbf {\bibinfo
  {volume} {14}},\ \bibinfo {pages} {1110 } (\bibinfo {year}
  {2015})}\BibitemShut {NoStop}%
\bibitem [{\citenamefont {Vromans}\ and\ \citenamefont
  {Giomi}(2015)}]{Vromans2015}%
  \BibitemOpen
  \bibfield  {author} {\bibinfo {author} {\bibfnamefont {A.~J.}\ \bibnamefont
  {Vromans}}\ and\ \bibinfo {author} {\bibfnamefont {L.}~\bibnamefont
  {Giomi}},\ }\href {\doibase 10.1039/C6SM01146B} {\bibfield  {journal}
  {\bibinfo  {journal} {Soft Matter}\ ,\ \bibinfo {pages} {1}} (\bibinfo {year}
  {2015})},\ \Eprint {http://arxiv.org/abs/1507.05588} {1507.05588}
  \BibitemShut {NoStop}%
\bibitem [{\citenamefont {Zhou}\ \emph {et~al.}(2017)\citenamefont {Zhou},
  \citenamefont {Shiyanovskii}, \citenamefont {Park},\ and\ \citenamefont
  {Lavrentovich}}]{Zhou2017}%
  \BibitemOpen
  \bibfield  {author} {\bibinfo {author} {\bibfnamefont {S.}~\bibnamefont
  {Zhou}}, \bibinfo {author} {\bibfnamefont {S.~V.}\ \bibnamefont
  {Shiyanovskii}}, \bibinfo {author} {\bibfnamefont {H.-S.}\ \bibnamefont
  {Park}}, \ and\ \bibinfo {author} {\bibfnamefont {O.~D.}\ \bibnamefont
  {Lavrentovich}},\ }\href {\doibase 10.1038/ncomms14974} {\bibfield  {journal}
  {\bibinfo  {journal} {Nat. Commun.}\ }\textbf {\bibinfo {volume} {8}},\
  \bibinfo {pages} {14974} (\bibinfo {year} {2017})}\BibitemShut {NoStop}%
\end{thebibliography}%


\begin{thebibliography}{56}%
\makeatletter
\providecommand \@ifxundefined [1]{%
 \@ifx{#1\undefined}
}%
\providecommand \@ifnum [1]{%
 \ifnum #1\expandafter \@firstoftwo
 \else \expandafter \@secondoftwo
 \fi
}%
\providecommand \@ifx [1]{%
 \ifx #1\expandafter \@firstoftwo
 \else \expandafter \@secondoftwo
 \fi
}%
\providecommand \natexlab [1]{#1}%
\providecommand \enquote  [1]{``#1''}%
\providecommand \bibnamefont  [1]{#1}%
\providecommand \bibfnamefont [1]{#1}%
\providecommand \citenamefont [1]{#1}%
\providecommand \href@noop [0]{\@secondoftwo}%
\providecommand \href [0]{\begingroup \@sanitize@url \@href}%
\providecommand \@href[1]{\@@startlink{#1}\@@href}%
\providecommand \@@href[1]{\endgroup#1\@@endlink}%
\providecommand \@sanitize@url [0]{\catcode `\\12\catcode `\$12\catcode
  `\&12\catcode `\#12\catcode `\^12\catcode `\_12\catcode `\%12\relax}%
\providecommand \@@startlink[1]{}%
\providecommand \@@endlink[0]{}%
\providecommand \url  [0]{\begingroup\@sanitize@url \@url }%
\providecommand \@url [1]{\endgroup\@href {#1}{\urlprefix }}%
\providecommand \urlprefix  [0]{URL }%
\providecommand \Eprint [0]{\href }%
\providecommand \doibase [0]{http://dx.doi.org/}%
\providecommand \selectlanguage [0]{\@gobble}%
\providecommand \bibinfo  [0]{\@secondoftwo}%
\providecommand \bibfield  [0]{\@secondoftwo}%
\providecommand \translation [1]{[#1]}%
\providecommand \BibitemOpen [0]{}%
\providecommand \bibitemStop [0]{}%
\providecommand \bibitemNoStop [0]{.\EOS\space}%
\providecommand \EOS [0]{\spacefactor3000\relax}%
\providecommand \BibitemShut  [1]{\csname bibitem#1\endcsname}%
\let\auto@bib@innerbib\@empty
\bibitem [{\citenamefont {Sanchez}\ \emph {et~al.}(2012)\citenamefont
  {Sanchez}, \citenamefont {Chen}, \citenamefont {Decamp}, \citenamefont
  {Heymann},\ and\ \citenamefont {Dogic}}]{Sanchez2012}%
  \BibitemOpen
  \bibfield  {author} {\bibinfo {author} {\bibfnamefont {T.}~\bibnamefont
  {Sanchez}}, \bibinfo {author} {\bibfnamefont {D.~T.~N.}\ \bibnamefont
  {Chen}}, \bibinfo {author} {\bibfnamefont {S.~J.}\ \bibnamefont {Decamp}},
  \bibinfo {author} {\bibfnamefont {M.}~\bibnamefont {Heymann}}, \ and\
  \bibinfo {author} {\bibfnamefont {Z.}~\bibnamefont {Dogic}},\ }\href
  {\doibase 10.1038/nature11591} {\bibfield  {journal} {\bibinfo  {journal}
  {Nature}\ }\textbf {\bibinfo {volume} {491}},\ \bibinfo {pages} {431–434}
  (\bibinfo {year} {2012})},\ \Eprint {http://arxiv.org/abs/1301.1122}
  {arXiv:1301.1122} \BibitemShut {NoStop}%
\bibitem [{\citenamefont {DeCamp}\ \emph {et~al.}(2015)\citenamefont {DeCamp},
  \citenamefont {Redner}, \citenamefont {Baskaran}, \citenamefont {Hagan},\
  and\ \citenamefont {Dogic}}]{DeCamp2015}%
  \BibitemOpen
  \bibfield  {author} {\bibinfo {author} {\bibfnamefont {S.~J.}\ \bibnamefont
  {DeCamp}}, \bibinfo {author} {\bibfnamefont {G.~S.}\ \bibnamefont {Redner}},
  \bibinfo {author} {\bibfnamefont {A.}~\bibnamefont {Baskaran}}, \bibinfo
  {author} {\bibfnamefont {M.~F.}\ \bibnamefont {Hagan}}, \ and\ \bibinfo
  {author} {\bibfnamefont {Z.}~\bibnamefont {Dogic}},\ }\href@noop {}
  {\bibfield  {journal} {\bibinfo  {journal} {Nat. Mater.}\ }\textbf {\bibinfo
  {volume} {14}},\ \bibinfo {pages} {1110 } (\bibinfo {year}
  {2015})}\BibitemShut {NoStop}%
\bibitem [{\citenamefont {Peng}\ \emph {et~al.}(2016)\citenamefont {Peng},
  \citenamefont {Turiv}, \citenamefont {Guo}, \citenamefont {Wei},\ and\
  \citenamefont {Lavrentovich}}]{peng2016command}%
  \BibitemOpen
  \bibfield  {author} {\bibinfo {author} {\bibfnamefont {C.}~\bibnamefont
  {Peng}}, \bibinfo {author} {\bibfnamefont {T.}~\bibnamefont {Turiv}},
  \bibinfo {author} {\bibfnamefont {Y.}~\bibnamefont {Guo}}, \bibinfo {author}
  {\bibfnamefont {Q.-H.}\ \bibnamefont {Wei}}, \ and\ \bibinfo {author}
  {\bibfnamefont {O.~D.}\ \bibnamefont {Lavrentovich}},\ }\href@noop {}
  {\bibfield  {journal} {\bibinfo  {journal} {Science}\ }\textbf {\bibinfo
  {volume} {354}},\ \bibinfo {pages} {882} (\bibinfo {year}
  {2016})}\BibitemShut {NoStop}%
\bibitem [{\citenamefont {Zhou}\ \emph {et~al.}(2014)\citenamefont {Zhou},
  \citenamefont {Sokolov}, \citenamefont {Lavrentovich},\ and\ \citenamefont
  {Aranson}}]{Zhou28012014}%
  \BibitemOpen
  \bibfield  {author} {\bibinfo {author} {\bibfnamefont {S.}~\bibnamefont
  {Zhou}}, \bibinfo {author} {\bibfnamefont {A.}~\bibnamefont {Sokolov}},
  \bibinfo {author} {\bibfnamefont {O.~D.}\ \bibnamefont {Lavrentovich}}, \
  and\ \bibinfo {author} {\bibfnamefont {I.~S.}\ \bibnamefont {Aranson}},\
  }\href {\doibase 10.1073/pnas.1321926111} {\bibfield  {journal} {\bibinfo
  {journal} {Proc. Natl. Acad. Sci. U. S. A.}\ }\textbf {\bibinfo {volume}
  {111}},\ \bibinfo {pages} {1265} (\bibinfo {year} {2014})},\ \Eprint
  {http://arxiv.org/abs/http://www.pnas.org/content/111/4/1265.full.pdf}
  {http://www.pnas.org/content/111/4/1265.full.pdf} \BibitemShut {NoStop}%
\bibitem [{\citenamefont {Narayan}\ \emph {et~al.}(2007)\citenamefont
  {Narayan}, \citenamefont {Ramaswamy},\ and\ \citenamefont
  {Menon}}]{Narayan2007}%
  \BibitemOpen
  \bibfield  {author} {\bibinfo {author} {\bibfnamefont {V.}~\bibnamefont
  {Narayan}}, \bibinfo {author} {\bibfnamefont {S.}~\bibnamefont {Ramaswamy}},
  \ and\ \bibinfo {author} {\bibfnamefont {N.}~\bibnamefont {Menon}},\ }\href
  {\doibase 10.1126/science.1140414} {\bibfield  {journal} {\bibinfo  {journal}
  {Science}\ }\textbf {\bibinfo {volume} {317}},\ \bibinfo {pages} {105}
  (\bibinfo {year} {2007})}\BibitemShut {NoStop}%
\bibitem [{\citenamefont {Narayan}\ \emph {et~al.}(2006)\citenamefont
  {Narayan}, \citenamefont {Menon},\ and\ \citenamefont
  {Ramaswamy}}]{narayan2006nonequilibrium}%
  \BibitemOpen
  \bibfield  {author} {\bibinfo {author} {\bibfnamefont {V.}~\bibnamefont
  {Narayan}}, \bibinfo {author} {\bibfnamefont {N.}~\bibnamefont {Menon}}, \
  and\ \bibinfo {author} {\bibfnamefont {S.}~\bibnamefont {Ramaswamy}},\
  }\href@noop {} {\bibfield  {journal} {\bibinfo  {journal} {J. Stat. Mech:
  Theory Exp.}\ }\textbf {\bibinfo {volume} {2006}},\ \bibinfo {pages} {P01005}
  (\bibinfo {year} {2006})}\BibitemShut {NoStop}%
\bibitem [{\citenamefont {Broer}(2010)}]{Broer2010}%
  \BibitemOpen
  \bibfield  {author} {\bibinfo {author} {\bibfnamefont {D.~J.}\ \bibnamefont
  {Broer}},\ }\href {\doibase 10.1038/nmat2617} {\bibfield  {journal} {\bibinfo
   {journal} {Nat. Mater.}\ }\textbf {\bibinfo {volume} {9}},\ \bibinfo {pages}
  {99} (\bibinfo {year} {2010})}\BibitemShut {NoStop}%
\bibitem [{\citenamefont {Lin}\ \emph {et~al.}(2011)\citenamefont {Lin},
  \citenamefont {Miller}, \citenamefont {Bertics}, \citenamefont {Murphy},
  \citenamefont {de~Pablo},\ and\ \citenamefont {Abbott}}]{Lin2011}%
  \BibitemOpen
  \bibfield  {author} {\bibinfo {author} {\bibfnamefont {I.-h.}\ \bibnamefont
  {Lin}}, \bibinfo {author} {\bibfnamefont {D.~S.}\ \bibnamefont {Miller}},
  \bibinfo {author} {\bibfnamefont {P.~J.}\ \bibnamefont {Bertics}}, \bibinfo
  {author} {\bibfnamefont {C.~J.}\ \bibnamefont {Murphy}}, \bibinfo {author}
  {\bibfnamefont {J.~J.}\ \bibnamefont {de~Pablo}}, \ and\ \bibinfo {author}
  {\bibfnamefont {N.~L.}\ \bibnamefont {Abbott}},\ }\href {\doibase
  10.1126/science.1195639} {\bibfield  {journal} {\bibinfo  {journal}
  {Science}\ }\textbf {\bibinfo {volume} {332}},\ \bibinfo {pages} {1297}
  (\bibinfo {year} {2011})}\BibitemShut {NoStop}%
\bibitem [{\citenamefont {Kleman}(1989)}]{Kleman1989}%
  \BibitemOpen
  \bibfield  {author} {\bibinfo {author} {\bibfnamefont {M.}~\bibnamefont
  {Kleman}},\ }\href {\doibase 10.1088/0034-4885/52/5/002} {\bibfield
  {journal} {\bibinfo  {journal} {Rep. Prog. Phys.}\ }\textbf {\bibinfo
  {volume} {52}},\ \bibinfo {pages} {555} (\bibinfo {year} {1989})}\BibitemShut
  {NoStop}%
\bibitem [{\citenamefont {Zhou}\ \emph {et~al.}(2017)\citenamefont {Zhou},
  \citenamefont {Shiyanovskii}, \citenamefont {Park},\ and\ \citenamefont
  {Lavrentovich}}]{Zhou2017}%
  \BibitemOpen
  \bibfield  {author} {\bibinfo {author} {\bibfnamefont {S.}~\bibnamefont
  {Zhou}}, \bibinfo {author} {\bibfnamefont {S.~V.}\ \bibnamefont
  {Shiyanovskii}}, \bibinfo {author} {\bibfnamefont {H.-S.}\ \bibnamefont
  {Park}}, \ and\ \bibinfo {author} {\bibfnamefont {O.~D.}\ \bibnamefont
  {Lavrentovich}},\ }\href {\doibase 10.1038/ncomms14974} {\bibfield  {journal}
  {\bibinfo  {journal} {Nat. Commun.}\ }\textbf {\bibinfo {volume} {8}},\
  \bibinfo {pages} {14974} (\bibinfo {year} {2017})}\BibitemShut {NoStop}%
\bibitem [{\citenamefont {Zhang}\ \emph {et~al.}(2018)\citenamefont {Zhang},
  \citenamefont {Kumar}, \citenamefont {Ross}, \citenamefont {Gardel},\ and\
  \citenamefont {de~Pablo}}]{Zhang2018}%
  \BibitemOpen
  \bibfield  {author} {\bibinfo {author} {\bibfnamefont {R.}~\bibnamefont
  {Zhang}}, \bibinfo {author} {\bibfnamefont {N.}~\bibnamefont {Kumar}},
  \bibinfo {author} {\bibfnamefont {J.~L.}\ \bibnamefont {Ross}}, \bibinfo
  {author} {\bibfnamefont {M.~L.}\ \bibnamefont {Gardel}}, \ and\ \bibinfo
  {author} {\bibfnamefont {J.~J.}\ \bibnamefont {de~Pablo}},\ }\href {\doibase
  10.1073/pnas.1713832115} {\bibfield  {journal} {\bibinfo  {journal} {Proc.
  Natl. Acad. Sci. U. S. A.}\ }\textbf {\bibinfo {volume} {115}},\ \bibinfo
  {pages} {E124} (\bibinfo {year} {2018})},\ \Eprint
  {http://arxiv.org/abs/http://www.pnas.org/content/115/2/E124.full.pdf}
  {http://www.pnas.org/content/115/2/E124.full.pdf} \BibitemShut {NoStop}%
\bibitem [{\citenamefont {Frank}(1958)}]{Frank1958}%
  \BibitemOpen
  \bibfield  {author} {\bibinfo {author} {\bibfnamefont {F.~C.}\ \bibnamefont
  {Frank}},\ }\href {\doibase 10.1039/df9582500019} {\bibfield  {journal}
  {\bibinfo  {journal} {Discuss. Faraday Soc.}\ }\textbf {\bibinfo {volume}
  {25}},\ \bibinfo {pages} {19} (\bibinfo {year} {1958})}\BibitemShut {NoStop}%
\bibitem [{\citenamefont {Oseen}(1933)}]{Oseen1933}%
  \BibitemOpen
  \bibfield  {author} {\bibinfo {author} {\bibfnamefont {C.~W.}\ \bibnamefont
  {Oseen}},\ }\href {\doibase 10.1039/TF9332900883} {\bibfield  {journal}
  {\bibinfo  {journal} {Trans. Faraday Soc.}\ }\textbf {\bibinfo {volume}
  {29}},\ \bibinfo {pages} {883} (\bibinfo {year} {1933})}\BibitemShut
  {NoStop}%
\bibitem [{\citenamefont {Simha}\ and\ \citenamefont
  {Ramaswamy}(2002)}]{Simha2002}%
  \BibitemOpen
  \bibfield  {author} {\bibinfo {author} {\bibfnamefont {R.~A.}\ \bibnamefont
  {Simha}}\ and\ \bibinfo {author} {\bibfnamefont {S.}~\bibnamefont
  {Ramaswamy}},\ }\href {\doibase 10.1103/PhysRevLett.89.058101} {\bibfield
  {journal} {\bibinfo  {journal} {Phys. Rev. Lett.}\ }\textbf {\bibinfo
  {volume} {89}},\ \bibinfo {pages} {058101} (\bibinfo {year}
  {2002})}\BibitemShut {NoStop}%
\bibitem [{\citenamefont {Marchetti}\ \emph {et~al.}(2013)\citenamefont
  {Marchetti}, \citenamefont {Joanny}, \citenamefont {Ramaswamy}, \citenamefont
  {Liverpool}, \citenamefont {Prost}, \citenamefont {Rao},\ and\ \citenamefont
  {Simha}}]{Marchetti2013}%
  \BibitemOpen
  \bibfield  {author} {\bibinfo {author} {\bibfnamefont {M.~C.}\ \bibnamefont
  {Marchetti}}, \bibinfo {author} {\bibfnamefont {J.~F.}\ \bibnamefont
  {Joanny}}, \bibinfo {author} {\bibfnamefont {S.}~\bibnamefont {Ramaswamy}},
  \bibinfo {author} {\bibfnamefont {T.~B.}\ \bibnamefont {Liverpool}}, \bibinfo
  {author} {\bibfnamefont {J.}~\bibnamefont {Prost}}, \bibinfo {author}
  {\bibfnamefont {M.}~\bibnamefont {Rao}}, \ and\ \bibinfo {author}
  {\bibfnamefont {R.~A.}\ \bibnamefont {Simha}},\ }\href {\doibase
  10.1103/RevModPhys.85.1143} {\bibfield  {journal} {\bibinfo  {journal} {Rev.
  Mod. Phys.}\ }\textbf {\bibinfo {volume} {85}},\ \bibinfo {pages} {1143}
  (\bibinfo {year} {2013})},\ \Eprint {http://arxiv.org/abs/1207.2929}
  {arXiv:1207.2929} \BibitemShut {NoStop}%
\bibitem [{\citenamefont {Ramaswamy}(2010)}]{Ramaswamy2010}%
  \BibitemOpen
  \bibfield  {author} {\bibinfo {author} {\bibfnamefont {S.}~\bibnamefont
  {Ramaswamy}},\ }\href {\doibase 10.1146/annurev-conmatphys-070909-104101}
  {\bibfield  {journal} {\bibinfo  {journal} {Annu. Rev. Condens. Matter
  Phys.}\ }\textbf {\bibinfo {volume} {1}},\ \bibinfo {pages} {323} (\bibinfo
  {year} {2010})},\ \Eprint {http://arxiv.org/abs/1004.1933} {arXiv:1004.1933}
  \BibitemShut {NoStop}%
\bibitem [{\citenamefont {Giomi}\ \emph {et~al.}(2013)\citenamefont {Giomi},
  \citenamefont {Bowick}, \citenamefont {Ma},\ and\ \citenamefont
  {Marchetti}}]{Giomi2013}%
  \BibitemOpen
  \bibfield  {author} {\bibinfo {author} {\bibfnamefont {L.}~\bibnamefont
  {Giomi}}, \bibinfo {author} {\bibfnamefont {M.~J.}\ \bibnamefont {Bowick}},
  \bibinfo {author} {\bibfnamefont {X.}~\bibnamefont {Ma}}, \ and\ \bibinfo
  {author} {\bibfnamefont {M.~C.}\ \bibnamefont {Marchetti}},\ }\href {\doibase
  10.1103/PhysRevLett.110.228101} {\bibfield  {journal} {\bibinfo  {journal}
  {Phys. Rev. Lett.}\ }\textbf {\bibinfo {volume} {110}},\ \bibinfo {pages}
  {228101} (\bibinfo {year} {2013})}\BibitemShut {NoStop}%
\bibitem [{\citenamefont {Giomi}\ \emph {et~al.}(2014)\citenamefont {Giomi},
  \citenamefont {Bowick}, \citenamefont {Mishra}, \citenamefont {Sknepnek},\
  and\ \citenamefont {{Cristina Marchetti}}}]{Giomi2014}%
  \BibitemOpen
  \bibfield  {author} {\bibinfo {author} {\bibfnamefont {L.}~\bibnamefont
  {Giomi}}, \bibinfo {author} {\bibfnamefont {M.~J.}\ \bibnamefont {Bowick}},
  \bibinfo {author} {\bibfnamefont {P.}~\bibnamefont {Mishra}}, \bibinfo
  {author} {\bibfnamefont {R.}~\bibnamefont {Sknepnek}}, \ and\ \bibinfo
  {author} {\bibfnamefont {M.}~\bibnamefont {{Cristina Marchetti}}},\ }\href
  {\doibase 10.1098/rsta.2013.0365} {\bibfield  {journal} {\bibinfo  {journal}
  {Phil. Trans. R. Soc. A}\ }\textbf {\bibinfo {volume} {372}},\ \bibinfo
  {pages} {20130365} (\bibinfo {year} {2014})}\BibitemShut {NoStop}%
\bibitem [{\citenamefont {Doostmohammadi}\ \emph {et~al.}(2017)\citenamefont
  {Doostmohammadi}, \citenamefont {Shendruk}, \citenamefont {Thijssen},\ and\
  \citenamefont {Yeomans}}]{Doostmohammadi2017}%
  \BibitemOpen
  \bibfield  {author} {\bibinfo {author} {\bibfnamefont {A.}~\bibnamefont
  {Doostmohammadi}}, \bibinfo {author} {\bibfnamefont {T.~N.}\ \bibnamefont
  {Shendruk}}, \bibinfo {author} {\bibfnamefont {K.}~\bibnamefont {Thijssen}},
  \ and\ \bibinfo {author} {\bibfnamefont {J.~M.}\ \bibnamefont {Yeomans}},\
  }\href {\doibase 10.1038/ncomms15326} {\bibfield  {journal} {\bibinfo
  {journal} {Nat. Comm.}\ }\textbf {\bibinfo {volume} {8}},\ \bibinfo {pages}
  {15326} (\bibinfo {year} {2017})},\ \Eprint {http://arxiv.org/abs/1607.01376}
  {arXiv:1607.01376} \BibitemShut {NoStop}%
\bibitem [{\citenamefont {Oza}\ and\ \citenamefont {Dunkel}(2016)}]{Oza2016}%
  \BibitemOpen
  \bibfield  {author} {\bibinfo {author} {\bibfnamefont {A.~U.}\ \bibnamefont
  {Oza}}\ and\ \bibinfo {author} {\bibfnamefont {J.}~\bibnamefont {Dunkel}},\
  }\href {\doibase 10.1088/1367-2630/18/9/093006} {\bibfield  {journal}
  {\bibinfo  {journal} {New J. Phys.}\ }\textbf {\bibinfo {volume} {18}},\
  \bibinfo {pages} {093006} (\bibinfo {year} {2016})},\ \Eprint
  {http://arxiv.org/abs/1507.01055} {arXiv:1507.01055} \BibitemShut {NoStop}%
\bibitem [{\citenamefont {Cortese}\ \emph {et~al.}(2018)\citenamefont
  {Cortese}, \citenamefont {Eggers},\ and\ \citenamefont
  {Liverpool}}]{Cortese2018}%
  \BibitemOpen
  \bibfield  {author} {\bibinfo {author} {\bibfnamefont {D.}~\bibnamefont
  {Cortese}}, \bibinfo {author} {\bibfnamefont {J.}~\bibnamefont {Eggers}}, \
  and\ \bibinfo {author} {\bibfnamefont {T.~B.}\ \bibnamefont {Liverpool}},\
  }\href {\doibase 10.1103/PhysRevE.97.022704} {\bibfield  {journal} {\bibinfo
  {journal} {Phys. Rev. E}\ }\textbf {\bibinfo {volume} {97}},\ \bibinfo
  {pages} {022704} (\bibinfo {year} {2018})}\BibitemShut {NoStop}%
\bibitem [{\citenamefont {Thampi}\ \emph {et~al.}(2014)\citenamefont {Thampi},
  \citenamefont {Golestanian},\ and\ \citenamefont {Yeomans}}]{Thampi2014}%
  \BibitemOpen
  \bibfield  {author} {\bibinfo {author} {\bibfnamefont {S.~P.}\ \bibnamefont
  {Thampi}}, \bibinfo {author} {\bibfnamefont {R.}~\bibnamefont {Golestanian}},
  \ and\ \bibinfo {author} {\bibfnamefont {J.~M.}\ \bibnamefont {Yeomans}},\
  }\href {\doibase 10.1209/0295-5075/105/18001} {\bibfield  {journal} {\bibinfo
   {journal} {EPL (Europhysics Letters)}\ }\textbf {\bibinfo {volume} {105}},\
  \bibinfo {pages} {18001} (\bibinfo {year} {2014})},\ \Eprint
  {http://arxiv.org/abs/1312.4836} {arXiv:1312.4836} \BibitemShut {NoStop}%
\bibitem [{\citenamefont {Giomi}(2015)}]{giomi2015geometry}%
  \BibitemOpen
  \bibfield  {author} {\bibinfo {author} {\bibfnamefont {L.}~\bibnamefont
  {Giomi}},\ }\href@noop {} {\bibfield  {journal} {\bibinfo  {journal} {Phys.
  Rev. X}\ }\textbf {\bibinfo {volume} {5}},\ \bibinfo {pages} {031003}
  (\bibinfo {year} {2015})}\BibitemShut {NoStop}%
\bibitem [{\citenamefont {Shendruk}\ \emph {et~al.}(2017)\citenamefont
  {Shendruk}, \citenamefont {Doostmohammadi}, \citenamefont {Thijssen},\ and\
  \citenamefont {Yeomans}}]{shendruk2017dancing}%
  \BibitemOpen
  \bibfield  {author} {\bibinfo {author} {\bibfnamefont {T.~N.}\ \bibnamefont
  {Shendruk}}, \bibinfo {author} {\bibfnamefont {A.}~\bibnamefont
  {Doostmohammadi}}, \bibinfo {author} {\bibfnamefont {K.}~\bibnamefont
  {Thijssen}}, \ and\ \bibinfo {author} {\bibfnamefont {J.~M.}\ \bibnamefont
  {Yeomans}},\ }\href@noop {} {\bibfield  {journal} {\bibinfo  {journal} {Soft
  Matter}\ }\textbf {\bibinfo {volume} {13}},\ \bibinfo {pages} {3853}
  (\bibinfo {year} {2017})}\BibitemShut {NoStop}%
\bibitem [{\citenamefont {Norton}\ \emph {et~al.}(2018)\citenamefont {Norton},
  \citenamefont {Baskaran}, \citenamefont {Opathalage}, \citenamefont
  {Langeslay}, \citenamefont {Fraden}, \citenamefont {Baskaran},\ and\
  \citenamefont {Hagan}}]{Norton2018}%
  \BibitemOpen
  \bibfield  {author} {\bibinfo {author} {\bibfnamefont {M.~M.}\ \bibnamefont
  {Norton}}, \bibinfo {author} {\bibfnamefont {A.}~\bibnamefont {Baskaran}},
  \bibinfo {author} {\bibfnamefont {A.}~\bibnamefont {Opathalage}}, \bibinfo
  {author} {\bibfnamefont {B.}~\bibnamefont {Langeslay}}, \bibinfo {author}
  {\bibfnamefont {S.}~\bibnamefont {Fraden}}, \bibinfo {author} {\bibfnamefont
  {A.}~\bibnamefont {Baskaran}}, \ and\ \bibinfo {author} {\bibfnamefont
  {M.~F.}\ \bibnamefont {Hagan}},\ }\href {\doibase 10.1103/PhysRevE.97.012702}
  {\bibfield  {journal} {\bibinfo  {journal} {Phys. Rev. E}\ }\textbf {\bibinfo
  {volume} {97}},\ \bibinfo {pages} {012702} (\bibinfo {year}
  {2018})}\BibitemShut {NoStop}%
\bibitem [{\citenamefont {Gao}\ \emph {et~al.}(2017)\citenamefont {Gao},
  \citenamefont {Betterton}, \citenamefont {Jhang},\ and\ \citenamefont
  {Shelley}}]{Gao2017}%
  \BibitemOpen
  \bibfield  {author} {\bibinfo {author} {\bibfnamefont {T.}~\bibnamefont
  {Gao}}, \bibinfo {author} {\bibfnamefont {M.~D.}\ \bibnamefont {Betterton}},
  \bibinfo {author} {\bibfnamefont {A.-S.}\ \bibnamefont {Jhang}}, \ and\
  \bibinfo {author} {\bibfnamefont {M.~J.}\ \bibnamefont {Shelley}},\ }\href
  {\doibase 10.1103/PhysRevFluids.2.093302} {\bibfield  {journal} {\bibinfo
  {journal} {Phys. Rev. Fluids}\ }\textbf {\bibinfo {volume} {2}},\ \bibinfo
  {pages} {093302} (\bibinfo {year} {2017})}\BibitemShut {NoStop}%
\bibitem [{\citenamefont {Keber}\ \emph {et~al.}(2014)\citenamefont {Keber},
  \citenamefont {Loiseau}, \citenamefont {Sanchez}, \citenamefont {DeCamp},
  \citenamefont {Giomi}, \citenamefont {Bowick}, \citenamefont {Marchetti},
  \citenamefont {Dogic},\ and\ \citenamefont {Bausch}}]{Keber2014}%
  \BibitemOpen
  \bibfield  {author} {\bibinfo {author} {\bibfnamefont {F.~C.}\ \bibnamefont
  {Keber}}, \bibinfo {author} {\bibfnamefont {E.}~\bibnamefont {Loiseau}},
  \bibinfo {author} {\bibfnamefont {T.}~\bibnamefont {Sanchez}}, \bibinfo
  {author} {\bibfnamefont {S.~J.}\ \bibnamefont {DeCamp}}, \bibinfo {author}
  {\bibfnamefont {L.}~\bibnamefont {Giomi}}, \bibinfo {author} {\bibfnamefont
  {M.~J.}\ \bibnamefont {Bowick}}, \bibinfo {author} {\bibfnamefont {M.~C.}\
  \bibnamefont {Marchetti}}, \bibinfo {author} {\bibfnamefont {Z.}~\bibnamefont
  {Dogic}}, \ and\ \bibinfo {author} {\bibfnamefont {A.~R.}\ \bibnamefont
  {Bausch}},\ }\href {\doibase 10.1126/science.1254784} {\bibfield  {journal}
  {\bibinfo  {journal} {Science}\ }\textbf {\bibinfo {volume} {345}},\ \bibinfo
  {pages} {1135} (\bibinfo {year} {2014})},\ \Eprint
  {http://arxiv.org/abs/15334406} {arXiv:15334406} \BibitemShut {NoStop}%
\bibitem [{\citenamefont {Zhang}\ \emph {et~al.}(2016)\citenamefont {Zhang},
  \citenamefont {Zhou}, \citenamefont {Rahimi},\ and\ \citenamefont
  {De~Pablo}}]{Zhang2016}%
  \BibitemOpen
  \bibfield  {author} {\bibinfo {author} {\bibfnamefont {R.}~\bibnamefont
  {Zhang}}, \bibinfo {author} {\bibfnamefont {Y.}~\bibnamefont {Zhou}},
  \bibinfo {author} {\bibfnamefont {M.}~\bibnamefont {Rahimi}}, \ and\ \bibinfo
  {author} {\bibfnamefont {J.~J.}\ \bibnamefont {De~Pablo}},\ }\href@noop {}
  {\bibfield  {journal} {\bibinfo  {journal} {Nat. Comm.}\ }\textbf {\bibinfo
  {volume} {7}},\ \bibinfo {pages} {13483} (\bibinfo {year}
  {2016})}\BibitemShut {NoStop}%
\bibitem [{\citenamefont {Ellis}\ \emph {et~al.}(2017)\citenamefont {Ellis},
  \citenamefont {Pearce}, \citenamefont {Chang}, \citenamefont {Goldsztein},
  \citenamefont {Giomi},\ and\ \citenamefont {Fernandez-Nieves}}]{Ellis2017}%
  \BibitemOpen
  \bibfield  {author} {\bibinfo {author} {\bibfnamefont {P.~W.}\ \bibnamefont
  {Ellis}}, \bibinfo {author} {\bibfnamefont {D.~J.~G.}\ \bibnamefont
  {Pearce}}, \bibinfo {author} {\bibfnamefont {Y.-W.}\ \bibnamefont {Chang}},
  \bibinfo {author} {\bibfnamefont {G.}~\bibnamefont {Goldsztein}}, \bibinfo
  {author} {\bibfnamefont {L.}~\bibnamefont {Giomi}}, \ and\ \bibinfo {author}
  {\bibfnamefont {A.}~\bibnamefont {Fernandez-Nieves}},\ }\href {\doibase
  10.1038/nphys4276
  https://www.nature.com/articles/nphys4276#supplementary-information}
  {\bibfield  {journal} {\bibinfo  {journal} {Nature Physics}\ }\textbf
  {\bibinfo {volume} {14}},\ \bibinfo {pages} {85} (\bibinfo {year}
  {2017})}\BibitemShut {NoStop}%
\bibitem [{\citenamefont {Alaimo}\ \emph {et~al.}(2017)\citenamefont {Alaimo},
  \citenamefont {K{\"{o}}hler},\ and\ \citenamefont {Voigt}}]{Alaimo2017}%
  \BibitemOpen
  \bibfield  {author} {\bibinfo {author} {\bibfnamefont {F.}~\bibnamefont
  {Alaimo}}, \bibinfo {author} {\bibfnamefont {C.}~\bibnamefont
  {K{\"{o}}hler}}, \ and\ \bibinfo {author} {\bibfnamefont {A.}~\bibnamefont
  {Voigt}},\ }\href {\doibase 10.1038/s41598-017-05612-6} {\bibfield  {journal}
  {\bibinfo  {journal} {Sci. Rep.}\ }\textbf {\bibinfo {volume} {7}},\ \bibinfo
  {pages} {5211} (\bibinfo {year} {2017})},\ \Eprint
  {http://arxiv.org/abs/1703.03707} {arXiv:1703.03707} \BibitemShut {NoStop}%
\bibitem [{\citenamefont {Gao}\ \emph {et~al.}(2015)\citenamefont {Gao},
  \citenamefont {Blackwell}, \citenamefont {Glaser}, \citenamefont
  {Betterton},\ and\ \citenamefont {Shelley}}]{Gao2015}%
  \BibitemOpen
  \bibfield  {author} {\bibinfo {author} {\bibfnamefont {T.}~\bibnamefont
  {Gao}}, \bibinfo {author} {\bibfnamefont {R.}~\bibnamefont {Blackwell}},
  \bibinfo {author} {\bibfnamefont {M.~A.}\ \bibnamefont {Glaser}}, \bibinfo
  {author} {\bibfnamefont {M.~D.}\ \bibnamefont {Betterton}}, \ and\ \bibinfo
  {author} {\bibfnamefont {M.~J.}\ \bibnamefont {Shelley}},\ }\href {\doibase
  10.1103/PhysRevLett.114.048101} {\bibfield  {journal} {\bibinfo  {journal}
  {Phys. Rev. Lett.}\ }\textbf {\bibinfo {volume} {114}},\ \bibinfo {pages}
  {048101} (\bibinfo {year} {2015})},\ \Eprint {http://arxiv.org/abs/1401.8059}
  {arXiv:1401.8059} \BibitemShut {NoStop}%
\bibitem [{\citenamefont {Shi}\ and\ \citenamefont {Ma}(2013)}]{Shi2013}%
  \BibitemOpen
  \bibfield  {author} {\bibinfo {author} {\bibfnamefont {X.-q.}\ \bibnamefont
  {Shi}}\ and\ \bibinfo {author} {\bibfnamefont {Y.-q.}\ \bibnamefont {Ma}},\
  }\href {\doibase 10.1038/ncomms4013} {\bibfield  {journal} {\bibinfo
  {journal} {Nat. Comm.}\ }\textbf {\bibinfo {volume} {4}},\ \bibinfo {pages}
  {3013} (\bibinfo {year} {2013})}\BibitemShut {NoStop}%
\bibitem [{\citenamefont {Sumino}\ \emph {et~al.}(2012)\citenamefont {Sumino},
  \citenamefont {Nagai}, \citenamefont {Shitaka}, \citenamefont {Tanaka},
  \citenamefont {Yoshikawa}, \citenamefont {Chat{\'e}},\ and\ \citenamefont
  {Oiwa}}]{sumino2012large}%
  \BibitemOpen
  \bibfield  {author} {\bibinfo {author} {\bibfnamefont {Y.}~\bibnamefont
  {Sumino}}, \bibinfo {author} {\bibfnamefont {K.~H.}\ \bibnamefont {Nagai}},
  \bibinfo {author} {\bibfnamefont {Y.}~\bibnamefont {Shitaka}}, \bibinfo
  {author} {\bibfnamefont {D.}~\bibnamefont {Tanaka}}, \bibinfo {author}
  {\bibfnamefont {K.}~\bibnamefont {Yoshikawa}}, \bibinfo {author}
  {\bibfnamefont {H.}~\bibnamefont {Chat{\'e}}}, \ and\ \bibinfo {author}
  {\bibfnamefont {K.}~\bibnamefont {Oiwa}},\ }\href@noop {} {\bibfield
  {journal} {\bibinfo  {journal} {Nature}\ }\textbf {\bibinfo {volume} {483}},\
  \bibinfo {pages} {448} (\bibinfo {year} {2012})}\BibitemShut {NoStop}%
\bibitem [{\citenamefont {Guillamat}\ \emph {et~al.}(2016)\citenamefont
  {Guillamat}, \citenamefont {Ignés-Mullol},\ and\ \citenamefont
  {Sagués}}]{Guillamat2016}%
  \BibitemOpen
  \bibfield  {author} {\bibinfo {author} {\bibfnamefont {P.}~\bibnamefont
  {Guillamat}}, \bibinfo {author} {\bibfnamefont {J.}~\bibnamefont
  {Ignés-Mullol}}, \ and\ \bibinfo {author} {\bibfnamefont {F.}~\bibnamefont
  {Sagués}},\ }\href {\doibase 10.1073/pnas.1600339113} {\bibfield  {journal}
  {\bibinfo  {journal} {Proc. Natl. Acad. Sci. U. S. A.}\ }\textbf {\bibinfo
  {volume} {113}},\ \bibinfo {pages} {5498} (\bibinfo {year} {2016})},\ \Eprint
  {http://arxiv.org/abs/http://www.pnas.org/content/113/20/5498.full.pdf}
  {http://www.pnas.org/content/113/20/5498.full.pdf} \BibitemShut {NoStop}%
\bibitem [{Met()}]{Metcalf2017}%
  \BibitemOpen
  \href@noop {} {}\bibinfo {note} {Image provided by Linnea Metcalf, Achini
  Opathalage and Zvonimir Dogc and used with their permission.}\BibitemShut
  {Stop}%
\bibitem [{Sup()}]{Supplement}%
  \BibitemOpen
  \href@noop {} {}\bibinfo {note} {Supplementary material to this paper is
  provided at [URL].}\BibitemShut {Stop}%
\bibitem [{\citenamefont {Kremer}\ and\ \citenamefont
  {Grest}(1990)}]{fene_1990}%
  \BibitemOpen
  \bibfield  {author} {\bibinfo {author} {\bibfnamefont {K.}~\bibnamefont
  {Kremer}}\ and\ \bibinfo {author} {\bibfnamefont {G.~S.}\ \bibnamefont
  {Grest}},\ }\href {\doibase 10.1063/1.458541} {\bibfield  {journal} {\bibinfo
   {journal} {J. Chem. Phys.}\ }\textbf {\bibinfo {volume} {92}},\ \bibinfo
  {pages} {5057} (\bibinfo {year} {1990})},\ \Eprint
  {http://arxiv.org/abs/http://dx.doi.org/10.1063/1.458541}
  {http://dx.doi.org/10.1063/1.458541} \BibitemShut {NoStop}%
\bibitem [{\citenamefont {Plimpton}(1995)}]{Plimpton1995}%
  \BibitemOpen
  \bibfield  {author} {\bibinfo {author} {\bibfnamefont {S.}~\bibnamefont
  {Plimpton}},\ }\href {\doibase 10.1006/jcph.1995.1039} {\bibfield  {journal}
  {\bibinfo  {journal} {J. Comput. Phys.}\ }\textbf {\bibinfo {volume} {117}},\
  \bibinfo {pages} {1} (\bibinfo {year} {1995})}\BibitemShut {NoStop}%
\bibitem [{\citenamefont {Prathyusha}\ \emph {et~al.}(2018)\citenamefont
  {Prathyusha}, \citenamefont {Henkes},\ and\ \citenamefont
  {Sknepnek}}]{Prathyusha2018}%
  \BibitemOpen
  \bibfield  {author} {\bibinfo {author} {\bibfnamefont {K.~R.}\ \bibnamefont
  {Prathyusha}}, \bibinfo {author} {\bibfnamefont {S.}~\bibnamefont {Henkes}},
  \ and\ \bibinfo {author} {\bibfnamefont {R.}~\bibnamefont {Sknepnek}},\
  }\href {\doibase 10.1103/PhysRevE.97.022606} {\bibfield  {journal} {\bibinfo
  {journal} {Phys. Rev. E}\ }\textbf {\bibinfo {volume} {97}},\ \bibinfo
  {pages} {022606} (\bibinfo {year} {2018})}\BibitemShut {NoStop}%
\bibitem [{\citenamefont {{Eppenga}}\ and\ \citenamefont
  {{Frenkel}}(1984)}]{Eppenga1984}%
  \BibitemOpen
  \bibfield  {author} {\bibinfo {author} {\bibfnamefont {R.}~\bibnamefont
  {{Eppenga}}}\ and\ \bibinfo {author} {\bibfnamefont {D.}~\bibnamefont
  {{Frenkel}}},\ }\href {\doibase 10.1080/00268978400101951} {\bibfield
  {journal} {\bibinfo  {journal} {Mol. Phys.}\ }\textbf {\bibinfo {volume}
  {52}},\ \bibinfo {pages} {1303} (\bibinfo {year} {1984})}\BibitemShut
  {NoStop}%
\bibitem [{\citenamefont {Baskaran}\ and\ \citenamefont
  {Marchetti}(2008)}]{Baskaran2008a}%
  \BibitemOpen
  \bibfield  {author} {\bibinfo {author} {\bibfnamefont {A.}~\bibnamefont
  {Baskaran}}\ and\ \bibinfo {author} {\bibfnamefont {M.~C.}\ \bibnamefont
  {Marchetti}},\ }\href {\doibase 10.1103/PhysRevLett.101.268101} {\bibfield
  {journal} {\bibinfo  {journal} {Phys. Rev. Lett}\ }\textbf {\bibinfo {volume}
  {101}},\ \bibinfo {pages} {268101} (\bibinfo {year} {2008})}\BibitemShut
  {NoStop}%
\bibitem [{\citenamefont {Putzig}\ \emph {et~al.}(2015)\citenamefont {Putzig},
  \citenamefont {Redner}, \citenamefont {Baskaran},\ and\ \citenamefont
  {Baskaran}}]{Putzig2015}%
  \BibitemOpen
  \bibfield  {author} {\bibinfo {author} {\bibfnamefont {E.}~\bibnamefont
  {Putzig}}, \bibinfo {author} {\bibfnamefont {G.~S.}\ \bibnamefont {Redner}},
  \bibinfo {author} {\bibfnamefont {A.}~\bibnamefont {Baskaran}}, \ and\
  \bibinfo {author} {\bibfnamefont {A.}~\bibnamefont {Baskaran}},\ }\href
  {\doibase 10.1039/C6SM00268D} {\bibfield  {journal} {\bibinfo  {journal}
  {Soft Matter}\ }\textbf {\bibinfo {volume} {12}},\ \bibinfo {pages} {1}
  (\bibinfo {year} {2015})},\ \Eprint {http://arxiv.org/abs/arXiv:1506.03501v1}
  {arXiv:arXiv:1506.03501v1} \BibitemShut {NoStop}%
\bibitem [{\citenamefont {qing Shi}\ \emph {et~al.}(2014)\citenamefont {qing
  Shi}, \citenamefont {Chaté},\ and\ \citenamefont {qiang Ma}}]{Shi2014}%
  \BibitemOpen
  \bibfield  {author} {\bibinfo {author} {\bibfnamefont {X.}~\bibnamefont {qing
  Shi}}, \bibinfo {author} {\bibfnamefont {H.}~\bibnamefont {Chaté}}, \ and\
  \bibinfo {author} {\bibfnamefont {Y.}~\bibnamefont {qiang Ma}},\ }\href
  {http://stacks.iop.org/1367-2630/16/i=3/a=035003} {\bibfield  {journal}
  {\bibinfo  {journal} {New J. Phys.}\ }\textbf {\bibinfo {volume} {16}},\
  \bibinfo {pages} {035003} (\bibinfo {year} {2014})}\BibitemShut {NoStop}%
\bibitem [{\citenamefont {Selinger}\ and\ \citenamefont
  {Bruinsma}(1991)}]{Selinger1991}%
  \BibitemOpen
  \bibfield  {author} {\bibinfo {author} {\bibfnamefont {J.~V.}\ \bibnamefont
  {Selinger}}\ and\ \bibinfo {author} {\bibfnamefont {R.~F.}\ \bibnamefont
  {Bruinsma}},\ }\href {\doibase 10.1103/PhysRevA.43.2910} {\bibfield
  {journal} {\bibinfo  {journal} {Phys. Rev. A}\ }\textbf {\bibinfo {volume}
  {43}},\ \bibinfo {pages} {2910} (\bibinfo {year} {1991})}\BibitemShut
  {NoStop}%
\bibitem [{\citenamefont {Dijkstra}\ and\ \citenamefont
  {Frenkel}(1995)}]{Dijkstra1995}%
  \BibitemOpen
  \bibfield  {author} {\bibinfo {author} {\bibfnamefont {M.}~\bibnamefont
  {Dijkstra}}\ and\ \bibinfo {author} {\bibfnamefont {D.}~\bibnamefont
  {Frenkel}},\ }\href {\doibase 10.1103/PhysRevE.51.5891} {\bibfield  {journal}
  {\bibinfo  {journal} {Phys. Rev. E}\ }\textbf {\bibinfo {volume} {51}},\
  \bibinfo {pages} {5891} (\bibinfo {year} {1995})}\BibitemShut {NoStop}%
\bibitem [{\citenamefont {Joshi}\ \emph {et~al.}(2014)\citenamefont {Joshi},
  \citenamefont {Whitmer}, \citenamefont {Guzm{\'{a}}n}, \citenamefont
  {Abbott},\ and\ \citenamefont {de~Pablo}}]{Joshi2014}%
  \BibitemOpen
  \bibfield  {author} {\bibinfo {author} {\bibfnamefont {A.~A.}\ \bibnamefont
  {Joshi}}, \bibinfo {author} {\bibfnamefont {J.~K.}\ \bibnamefont {Whitmer}},
  \bibinfo {author} {\bibfnamefont {O.}~\bibnamefont {Guzm{\'{a}}n}}, \bibinfo
  {author} {\bibfnamefont {N.~L.}\ \bibnamefont {Abbott}}, \ and\ \bibinfo
  {author} {\bibfnamefont {J.~J.}\ \bibnamefont {de~Pablo}},\ }\href {\doibase
  10.1039/C3SM51919H} {\bibfield  {journal} {\bibinfo  {journal} {Soft Matter}\
  }\textbf {\bibinfo {volume} {10}},\ \bibinfo {pages} {882} (\bibinfo {year}
  {2014})}\BibitemShut {NoStop}%
\bibitem [{Note1()}]{Note1}%
  \BibitemOpen
  \bibinfo {note} {Note that this 2D system with soft-core interactions should
  exhibit quasi-long-range order in the nematic phase ~\cite
  {Straley1971,FrenkelEppenga1985,Kosterlitz1973}, and thus $S$ decays with
  system size. Since we are interested in the degree of order on the scale of
  typical bend and splay fluctuations, we report $S$ values measured within
  regions of size $10\times 10 \sigma ^2$ (see \cite
  {Supplement}).}\BibitemShut {Stop}%
\bibitem [{\citenamefont {Mishra}\ and\ \citenamefont
  {Ramaswamy}(2006)}]{Mishra2006}%
  \BibitemOpen
  \bibfield  {author} {\bibinfo {author} {\bibfnamefont {S.}~\bibnamefont
  {Mishra}}\ and\ \bibinfo {author} {\bibfnamefont {S.}~\bibnamefont
  {Ramaswamy}},\ }\href {\doibase 10.1103/PhysRevLett.97.090602} {\bibfield
  {journal} {\bibinfo  {journal} {Phys. Rev. Lett.}\ }\textbf {\bibinfo
  {volume} {97}},\ \bibinfo {pages} {090602} (\bibinfo {year}
  {2006})}\BibitemShut {NoStop}%
\bibitem [{\citenamefont {Ngo}\ \emph {et~al.}(2014)\citenamefont {Ngo},
  \citenamefont {Peshkov}, \citenamefont {Aranson}, \citenamefont {Bertin},
  \citenamefont {Ginelli},\ and\ \citenamefont {Chat{\'{e}}}}]{Ngo2014}%
  \BibitemOpen
  \bibfield  {author} {\bibinfo {author} {\bibfnamefont {S.}~\bibnamefont
  {Ngo}}, \bibinfo {author} {\bibfnamefont {A.}~\bibnamefont {Peshkov}},
  \bibinfo {author} {\bibfnamefont {I.~S.}\ \bibnamefont {Aranson}}, \bibinfo
  {author} {\bibfnamefont {E.}~\bibnamefont {Bertin}}, \bibinfo {author}
  {\bibfnamefont {F.}~\bibnamefont {Ginelli}}, \ and\ \bibinfo {author}
  {\bibfnamefont {H.}~\bibnamefont {Chat{\'{e}}}},\ }\href {\doibase
  10.1103/PhysRevLett.113.038302} {\bibfield  {journal} {\bibinfo  {journal}
  {Phys. Rev. Lett.}\ }\textbf {\bibinfo {volume} {113}},\ \bibinfo {pages}
  {038302} (\bibinfo {year} {2014})},\ \Eprint {http://arxiv.org/abs/1312.1076}
  {arXiv:1312.1076} \BibitemShut {NoStop}%
\bibitem [{\citenamefont {Putzig}\ and\ \citenamefont
  {Baskaran}(2014)}]{Putzig2014}%
  \BibitemOpen
  \bibfield  {author} {\bibinfo {author} {\bibfnamefont {E.}~\bibnamefont
  {Putzig}}\ and\ \bibinfo {author} {\bibfnamefont {A.}~\bibnamefont
  {Baskaran}},\ }\href {\doibase 10.1103/PhysRevE.90.042304} {\bibfield
  {journal} {\bibinfo  {journal} {Phys. Rev. E}\ }\textbf {\bibinfo {volume}
  {90}},\ \bibinfo {pages} {042304} (\bibinfo {year} {2014})},\ \Eprint
  {http://arxiv.org/abs/1057984} {1057984} \BibitemShut {NoStop}%
\bibitem [{\citenamefont {Ramaswamy}\ \emph {et~al.}(2003)\citenamefont
  {Ramaswamy}, \citenamefont {Simha},\ and\ \citenamefont
  {Toner}}]{Ramaswamy2003}%
  \BibitemOpen
  \bibfield  {author} {\bibinfo {author} {\bibfnamefont {S.}~\bibnamefont
  {Ramaswamy}}, \bibinfo {author} {\bibfnamefont {R.~A.}\ \bibnamefont
  {Simha}}, \ and\ \bibinfo {author} {\bibfnamefont {J.}~\bibnamefont
  {Toner}},\ }\href {\doibase 10.1209/epl/i2003-00346-7} {\bibfield  {journal}
  {\bibinfo  {journal} {Europhys. Lett.}\ }\textbf {\bibinfo {volume} {62}},\
  \bibinfo {pages} {196} (\bibinfo {year} {2003})}\BibitemShut {NoStop}%
\bibitem [{\citenamefont {Chat\'{e}}\ \emph {et~al.}(2006)\citenamefont
  {Chat\'{e}}, \citenamefont {Ginelli},\ and\ \citenamefont
  {Montagne}}]{Chate2006}%
  \BibitemOpen
  \bibfield  {author} {\bibinfo {author} {\bibfnamefont {H.}~\bibnamefont
  {Chat\'{e}}}, \bibinfo {author} {\bibfnamefont {F.}~\bibnamefont {Ginelli}},
  \ and\ \bibinfo {author} {\bibfnamefont {R.}~\bibnamefont {Montagne}},\
  }\href {\doibase 10.1103/PhysRevLett.96.180602} {\bibfield  {journal}
  {\bibinfo  {journal} {Phys. Rev. Lett}\ }\textbf {\bibinfo {volume} {96}},\
  \bibinfo {pages} {180602} (\bibinfo {year} {2006})}\BibitemShut {NoStop}%
\bibitem [{\citenamefont {Zhang}\ \emph {et~al.}(2010)\citenamefont {Zhang},
  \citenamefont {Be{\textquoteright}er}, \citenamefont {Florin},\ and\
  \citenamefont {Swinney}}]{Zhang2010}%
  \BibitemOpen
  \bibfield  {author} {\bibinfo {author} {\bibfnamefont {H.~P.}\ \bibnamefont
  {Zhang}}, \bibinfo {author} {\bibfnamefont {A.}~\bibnamefont
  {Be{\textquoteright}er}}, \bibinfo {author} {\bibfnamefont {E.-L.}\
  \bibnamefont {Florin}}, \ and\ \bibinfo {author} {\bibfnamefont {H.~L.}\
  \bibnamefont {Swinney}},\ }\href {\doibase 10.1073/pnas.1001651107}
  {\bibfield  {journal} {\bibinfo  {journal} {Proc. Natl. Acad. Sci. U. S. A.}\
  }\textbf {\bibinfo {volume} {107}},\ \bibinfo {pages} {13626} (\bibinfo
  {year} {2010})},\ \Eprint
  {http://arxiv.org/abs/http://www.pnas.org/content/107/31/13626.full.pdf}
  {http://www.pnas.org/content/107/31/13626.full.pdf} \BibitemShut {NoStop}%
\bibitem [{\citenamefont {Straley}(1971)}]{Straley1971}%
  \BibitemOpen
  \bibfield  {author} {\bibinfo {author} {\bibfnamefont {J.~P.}\ \bibnamefont
  {Straley}},\ }\href {\doibase 10.1103/PhysRevA.4.675} {\bibfield  {journal}
  {\bibinfo  {journal} {Phys. Rev. A}\ }\textbf {\bibinfo {volume} {4}},\
  \bibinfo {pages} {675} (\bibinfo {year} {1971})}\BibitemShut {NoStop}%
\bibitem [{\citenamefont {Frenkel}\ and\ \citenamefont
  {Eppenga}(1985)}]{FrenkelEppenga1985}%
  \BibitemOpen
  \bibfield  {author} {\bibinfo {author} {\bibfnamefont {D.}~\bibnamefont
  {Frenkel}}\ and\ \bibinfo {author} {\bibfnamefont {R.}~\bibnamefont
  {Eppenga}},\ }\href {\doibase 10.1103/PhysRevA.31.1776} {\bibfield  {journal}
  {\bibinfo  {journal} {Phys. Rev. A}\ }\textbf {\bibinfo {volume} {31}},\
  \bibinfo {pages} {1776} (\bibinfo {year} {1985})}\BibitemShut {NoStop}%
\bibitem [{\citenamefont {Kosterlitz}\ and\ \citenamefont
  {Thouless}(1973)}]{Kosterlitz1973}%
  \BibitemOpen
  \bibfield  {author} {\bibinfo {author} {\bibfnamefont {J.~M.}\ \bibnamefont
  {Kosterlitz}}\ and\ \bibinfo {author} {\bibfnamefont {D.~J.}\ \bibnamefont
  {Thouless}},\ }\href {http://stacks.iop.org/0022-3719/6/i=7/a=010} {\bibfield
   {journal} {\bibinfo  {journal} {J. Phys. C: Solid State Phys.}\ }\textbf
  {\bibinfo {volume} {6}},\ \bibinfo {pages} {1181} (\bibinfo {year}
  {1973})}\BibitemShut {NoStop}%
\end{thebibliography}

%

\end{document}



\title{Supplementary material: The interplay of activity and filament flexibility determines the emergent properties of active nematics}

\author{Abhijeet Joshi}
\affiliation{Martin Fisher school of physics, Brandeis University, Waltham, MA 02453, USA}
\author{Elias Putzig}
\affiliation{Martin Fisher school of physics, Brandeis University, Waltham, MA 02453, USA}
\author{Aparna Baskaran}
\affiliation{Martin Fisher school of physics, Brandeis University, Waltham, MA 02453, USA}
\author{Michael F. Hagan}
\affiliation{Martin Fisher school of physics, Brandeis University, Waltham, MA 02453, USA}

\date{\today}

\maketitle

\setcounter{equation}{0}
\renewcommand{\theequation}{S\arabic{equation}}
\setcounter{figure}{0}
\renewcommand{\thefigure}{S\arabic{figure}}
\setcounter{table}{0}
\renewcommand{\thetable}{S\arabic{table}}
\setcounter{section}{0}
\renewcommand{\thesection}{S\arabic{section}}

\tableofcontents


\section{Simulation movies}
Animations of simulation trajectories (showing $1/16$ of the simulation box) are provided for the following parameter sets, with $\kfene=300$ and $\tau_1=0.2$ in all cases:
\begin{itemize}
  \item fa5\_k500\_t0p2.mp4: \quad  $\ktp=20: \kappa=500, \fa=5$
  \item fa5\_k2500\_t0p2.mp4: \quad $\ktp=100: \kappa=2500, \fa=5$
  \item fa10\_k200\_t0p2.mp4: \quad $\ktp=2: \kappa=200, \fa=10$
  \item fa10\_k2500\_t0p2.mp4: \quad $\ktp=25: \kappa=2500, \fa=10$
  \item fa30\_k10000\_t0p2.mp4: \quad $\ktp=11: \kappa=10^4, \fa=30$
\end{itemize}


In the videos, white arrows indicate positions and orientations of $\phd$ defects and white dots indicate positions of $\mhd$ defects. Filament beads are colored according to the orientations of the local tangent vector.


\section{Model and simulation details}
\label{sec:model}

\textit{Interaction potentials:}
We simulate the dynamics for the system of active filaments according to the following Langevin equation for each filament $\alpha$ and bead $i$ (with filaments indexed by Greek letters, $\alpha=1\ldots N$, and beads within a filament indexed in Roman, $i=1\ldots M$)
\begin{equation}
m\ddot{\mathbf{r}}_{\alpha,i} = \bfa_{\alpha,i} -\gamma\dot{\mathbf{r}}_{\alpha,i} - \nabla_{\mathbf{r}_{\alpha,i}}U + \mathbf{R}_{\alpha,i}(t).
\label{eq:Langevin}
\end{equation}
with $m$ as the bead mass, $\mathbf{r}_{\alpha,i}$ as the bead position,  $\fa$ as the active force, $U$ as the interaction potential which gives rise to the conservative forces, $\gamma$ as the friction coefficient providing the damping and $\mathbf{R}_{\alpha,i}(t)$ as a delta-correlated
thermal noise with zero mean and variance $\langle \mathbf{R}_{\alpha,i}(t)\cdot\mathbf{R}_{\beta,j}(t')\rangle=4\gamma \kt \delta_{\alpha,\beta} \delta_{ij}\delta(t-t')$.

The interaction potential includes three contributions which respectively account for non-bonded interactions between all bead pairs, stretching of each bond, and the angle potential between each pair of neighboring bonds:
\begin{align}
U(\{\mathbf{r}_{\alpha,i}\})=& \frac{1}{2}\sum_{\alpha,\beta=1}^{N}\sum_{i,j=1}^{M}\left(1-\delta_{\alpha,\beta}\delta_{i,j}\right) \Unb(\left| \mr_{\beta,j}-\mr_{\alpha,i}\right|) \nonumber \\
 + & \sum_{\alpha=1}^{N} \sum_{i=2}^{M} \Ustretch(\left|\mr_{\alpha,i}-\mr_{\alpha,i-1}\right|) \nonumber \\
 + &\sum_{\alpha=1}^{N} \sum_{i=3}^{M}\Uangle(\theta_{\alpha,i,i-1,i-2})
\label{eq:U}
\end{align}
with $\theta_{\alpha,i,j,k}$ the angle made by the bead triplet $\{i,j,k\}$ on filament $\alpha$. The non-bonded interactions account for steric repulsion and are
modeled with the Weeks-Chandler-Anderson potential\cite{Weeks1971}
\begin{align}
\Unb(r)=4\varepsilon\left((\sigma/r)^{12}-(\sigma/r)^6+1/4\right)\Theta(2^{1/6}\sigma-r)
\label{eq:Unb}
\end{align}
with $\varepsilon$ controlling the strength of steric repulsion and $\Theta(x)$ the Heaviside function specifying the cutoff. Bond stretching is controlled by a FENE potential \cite{fene_1990}
\begin{align}
\Ustretch(r)=-1/2 k_\text{b} R_0^2\ln\left(1-\left({r}/{R_0}\right)^2 \right)
\label{eq:Ustretch}
\end{align}
with bond strength $k_\text{b}$ and maximum bond length $R_0$. The angle potential is given by
\begin{align}
\Uangle(\theta)=\kappa\left(\theta -\pi \right)^2
\label{eq:Uangle}
\end{align}
where $\kappa$ is the filament bending modulus.


Finally, activity is modeled as a propulsive force on every bead directed along the filament tangent toward its head. To render the activity nematic, the head and tail of each filament are switched at stochastic intervals so that the force direction on each bead rotates by $180$ degrees. In particular,
the active force has the form, $\bfa_{\alpha,i}(t)=\eta_{\alpha}(t) \fa \mathbf{t}_{\alpha,i}$, where $\fa$ parameterizes the activity strength, and $\eta_{\alpha}(t)$ is a stochastic variable associated with filament $\alpha$ that changes its values between 1 and $-1$  on Poisson distributed intervals with mean $\tau_1$, so that  $\tau_1$ controls the reversal frequency.
 The local tangent vector $\mathbf{t}_{\alpha,i}$ along the filament contour at a bead $i$ is calculated as
\[
    \mathbf{t}_{\alpha,i} =
\begin{cases}
    \frac{\mr_{\alpha,i+1}-\mr_{\alpha,i-1}}{|\mr_{\alpha,i+1}-\mr_{\alpha,i-1}|}, & \text{for } i= 2,...,M-1\\
    \frac{\mr_{\alpha,2}-\mr_{\alpha,1}}{|\mr_{\alpha,2} -\mr_{\alpha,1}|},      & \text{for } i=1\\
    \frac{\mr_{\alpha,M}-\mr_{\alpha,M-1}}{|\mr_{\alpha,M}-\mr_{\alpha,M-1}|} & \text{for } i = M.
\end{cases}
\]

\textit{Simulations and parameter values:} We set mass of the each bead to $m=1$ and the damping coefficient to $\gamma=2$. With these parameters inertia is non-negligible. In the future, we plan to systematically investigate the effect of damping.    We report lengths and energies in units of the bead diameter $\sigma=1$ and \rev{the thermal energy at a reference state, $\ktRef$. Time is measured in units of $\tau=\sqrt{m\sigma^2/\ktRef}$. Eqs.~(\ref{eq:Langevin}) were integrated with time
step $\delta t=10^{-3}\tau$ using LAMMPS~\cite{Plimpton1995}, with an in-house modification
to include the active propulsion force. We set the repulsion parameter $\varepsilon=\ktRef$ (the results are insensitive to the value of $\varepsilon$ provided it is sufficiently high to avoid filament overlap)}. We performed simulations in a $A=840\times840\sigma^2$ periodic simulation box, with $N=19404$ filaments each with $M=20$ beads, so that the packing fraction $\phi=MN\sigma^2/A \approx 0.55$.

In the FENE bond potential, $R_0$ is set to $1.5\sigma$ and   $k_\text{b}=300\ktRef/\sigma^2$ for the simulations in the main text (see next paragraph). The temperature is set to  $T=3.0\ktRef$. These parameters lead to a mean bond length of $b\approx0.84\sigma$, which ensures
that filaments are non-penetrable for the parameter space explored in this work.
Finally, $L=(M-1)b$ is the mean filament length. In semiflexible limit, the
stiffness $\kappa$ in the discrete model is related to the
continuum bending modulus $\tilde{\kappa}$ as $\kappa\approx\tilde{\kappa}/2b$, and
thus the persistence length is given by $\lp=2 b \kappa/\kt$.

\rev{We performed two sets of simulations. Initially, we set the FENE bond strength $\kfene=30 \ktRef/\sigma^2$ and $\tau_1=\tau$, thus maintaining the reversal frequency of active propulsion to a fixed value  while varying  filament stiffness,
$\kappa \in [100,2500]\ktRef$ and the magnitude of the active force.  However, for these parameters we discovered that interpenetration of filaments becomes possible at large propulsion velocities, thus limiting the simulations to $\fa\le10$. To enable investigating higher activity values, we therefore performed a second set of simulations (those reported in the main text) with higher FENE bond strength, $\kfene=300 \ktRef/\sigma^2$ and shorter reversal timescale $\tau_1=0.2$, with $\kappa \in [100,10000]\ktRef$. This enables simulating activity values up to $\fa\le30$. }

\rev{ The shorter reversal timescale was needed in the new set of simulations because when the product $\fa\tau$ exceeds a characteristic collision length scale the self-propulsion becomes polar in nature and model is no longer a good description of an active nematic. In particular, the filaments behave as polar self-propelled rods, as evidenced by the formation of polar clusters \cite{Peruani2006,McCandlish2012,Weitz2015,Ginelli2010,Peruani2012}. Interestingly, increasing $\kfene$ increases the rate of defect formation and decreases the threshold value of $\fa\tau$ above which phase separation occurs.  We monitored the existence of phase separation by measuring local densities within simulation boxes (see below). We found no significant phase separation (except on short scales, see below) or formation of polar clusters for any of the simulations described here, indicating the systems were in the nematic regime. Within the nematic regime, increasing $\kfene$ does not qualitatively change results or scaling relations, although it does quantitatively shift properties such as the defect density. The scaling of defect density with $\kappa$ and $\fa$ for the simulations with $\kfene=30$ is shown in Fig.~\ref{fig:old_fig_3}.
}

\rev{In our simulations we have fixed the filament length at $M=20$ beads. Exploratory simulations showed that varying $M$ does not change the scaling of observables with $\ktp$, although it shifts properties such as the defect density since the total active force per filament goes as $\fa M$. The maximum persistence length above which scaling laws fail is also proportional to filament length: $\ktp^\text{max} = \kt M/ 2 b$ }

\textit{Initialization:}
We initialized the system by first placing the filaments in completely extended configurations (all bonds parallel and all bond lengths set to $b$),
on a rectangular lattice aligned with the simulation box, with filaments oriented along one of the lattice vectors.
We then allowed this initial configuration to relax by simulating for $10^4\tau$, before performing production runs of $\sim4\times10^4\tau$.
\rev{As noted in the main text, the equilibration time of $10^4\tau$ was chosen based on the fact that in all simulated systems the defect density had reached steady state by this time. We also confirmed that other observables of interest have reached steady state at this time.}
In the limit of high stiffness or low activity, we found that defect nucleation from the unphysically crystalline initial condition did not occur on computationally accessible timescales, thus prohibiting relaxation. In these cases, we used an alternative initial configuration, with filaments arranged into four rectangular
lattices, each placed in one quadrant of the simulation box such that adjacent (non-diagonal) lattices are orthogonal.

\textit{Analysis:} To analyze our simulation trajectories within the framework of liquid crystal theory, we
calculated the local nematic tensor as $Q_{\alpha\beta}(r_{ij})= \sum_{k, |r_k -r_{ij}| < 5\sigma}(\hat{t}_{k\alpha}\hat{t}_{k\beta}-\frac{1}{2}{\delta_{\alpha\beta}})$
and the density field $\rho$ on a $1000 \times 1000$ grid, where, $r_{ij}$ is the position
of the grid point and $r_k$ and $t_k$ are the positions and tangent vectors of all segments.
 We then evaluated the local nematic order parameter, $S$ and director $\hat{n}$  as the largest
 eigenvalue of $Q$ and corresponding eigenvector.  We identified defects in the director
 field and their topological charges using the procedure described in section \ref{measuring_defect_shape} below.  To
 compare the magnitudes of splay and bend deformations in our active systems to those
 that occur at equilibrium, we calculated elastic constants of our system at
 equilibrium ($\fa=0$) using the free energy  perturbation technique
 proposed by Joshi et al.~\cite{Joshi2014}. We performed these equilibrium
 calculations on a box of size $64 \times 64 \sigma^2$.

\section{Additional figures on scaling of active nematic characteristics with $\ktp$}

\begin{figure}[hbt]
\includegraphics[width=0.7\linewidth]{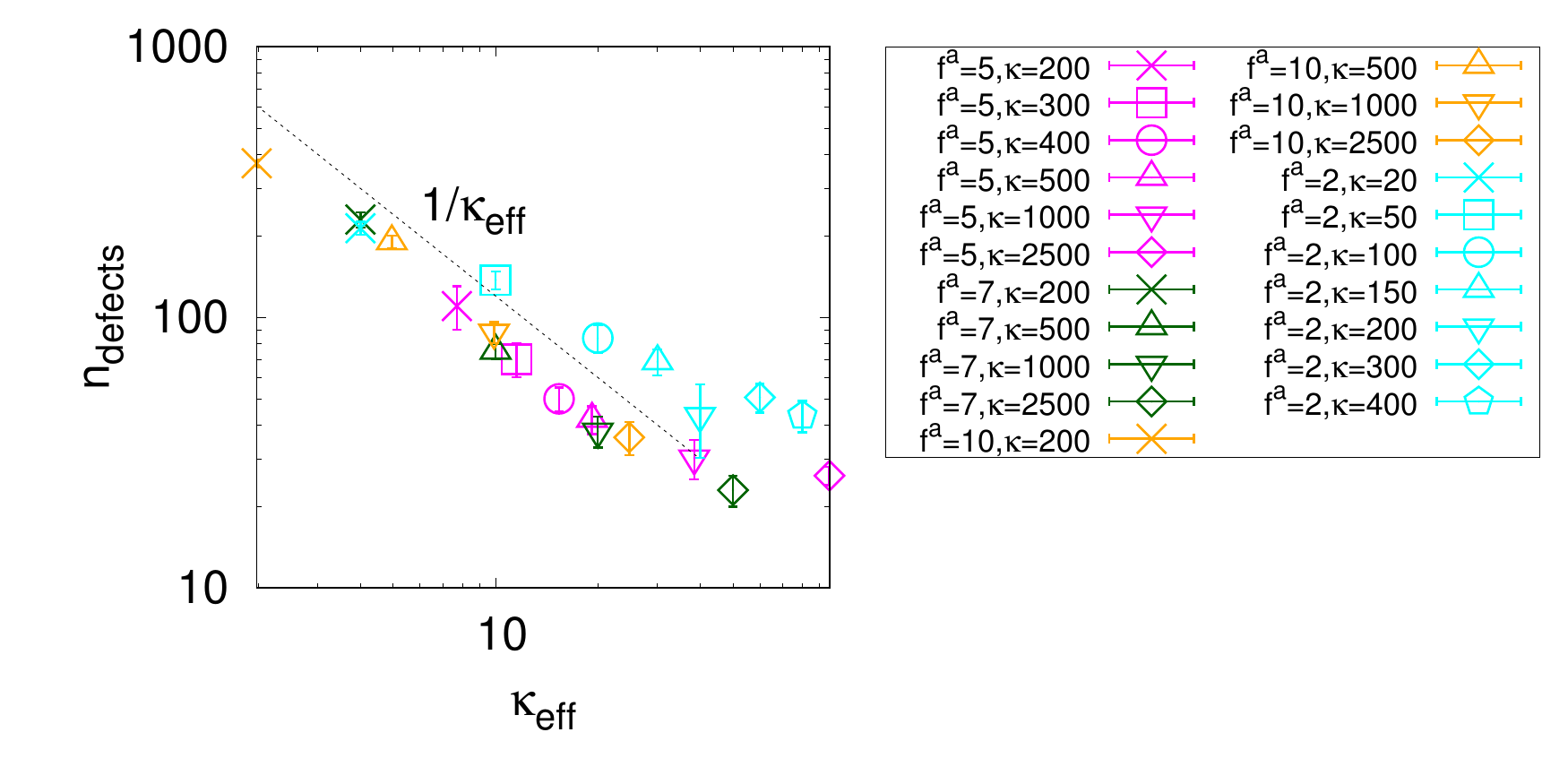}
\caption{\label{fig:old_fig_3} \rev{Number of defects as a function of $\ktp$ for the alternative dataset, with FENE bond strength $\kfene=30$ and active reversal period, $\tau_1=1$. Notice that we observe data collapse for all parameters with $\fa\ge5$, but not for the lowest activity $\fa=2$ (light blue symbols), when the system begins to lose nematic order (see Fig.~\ref{fig:kii}c below) and the active force strength becomes comparable to thermal forces.
}}
\end{figure}

\begin{figure}[hbt]
\includegraphics[width=\linewidth]{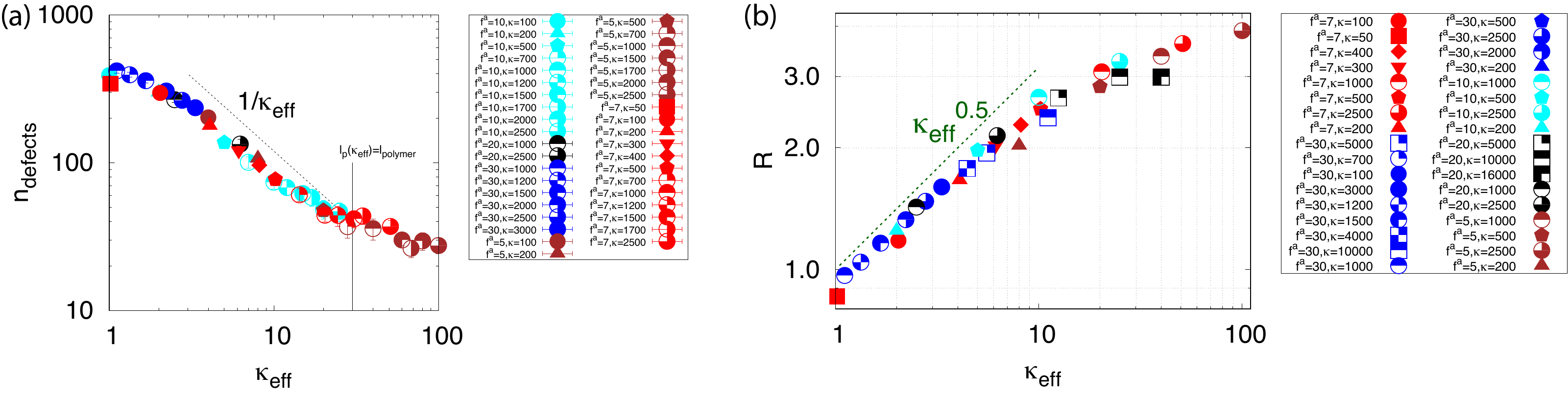}
\caption{\label{fig:all_data} \rev{Specification of all data points considered for the main data set ($\kfene=300$ and $\tau_1=0.2$)  \textbf{(a)} Defect density as a function of $\ktp$ for indicated values of $\fa$ and $\kappa$.  \textbf{(b)} Ratio of splay/bend ($R$, defined in the main text) as a function of $\ktp$ for indicated values of $\fa$ and $\kappa$.}
}
\end{figure}

We present data on defect density from the alternative data set described in section~\ref{sec:model} in Fig.~\ref{fig:old_fig_3}, and an explicit listing of parameter values for Fig. 3 in the main text in Fig.~\ref{fig:all_data}.

\color{black}
\section{Estimating the individual filament persistence length}

In this section we describe estimates of the effective persistence length measured from the tangent fluctuations of individual filaments. We performed these measurements both on individual filaments within a bulk active nematic, and isolated individual filaments to distinguish single-chain and multi-chain effects on the effective persistence length.

In a continuum limit, the total bending energy of a semiflexible filament it is well approximated by the wormlike chain model \cite{Doi1988},
\begin{equation}
 H_\text{bend} = \frac{\tilde{\kappa}}{2}\int_0^L \left(\frac{d\theta}{ds} \right)^2ds
 \label{eq:hbend_cont}
\end{equation}
where the integration is over the filament contour length, $L$, parameterized by $s$, $\tilde{\kappa}$ is the continuum bending modulus, and $\theta(s)$ is the tangent angle along the contour.

For a normal-mode analysis of the bending excitations
we performed a Fourier decomposition of the
tangential angle $\theta(s)$ assuming general boundary conditions (since a filament in bulk need not be force-free at its ends):
\begin{equation}
 \theta(s) = \sum_q (a_q \cos(qs) + b_q \cos(qs))
\label{eq:thetaq}
\end{equation}
where $q = n\pi/L$ $(n = 1, 2, 3 ... )$ is the wave vector, with corresponding wavelength $\lambda = \pi/q$.

\begin{figure}[h]
\vspace{1cm}
\includegraphics[width=0.7\linewidth]{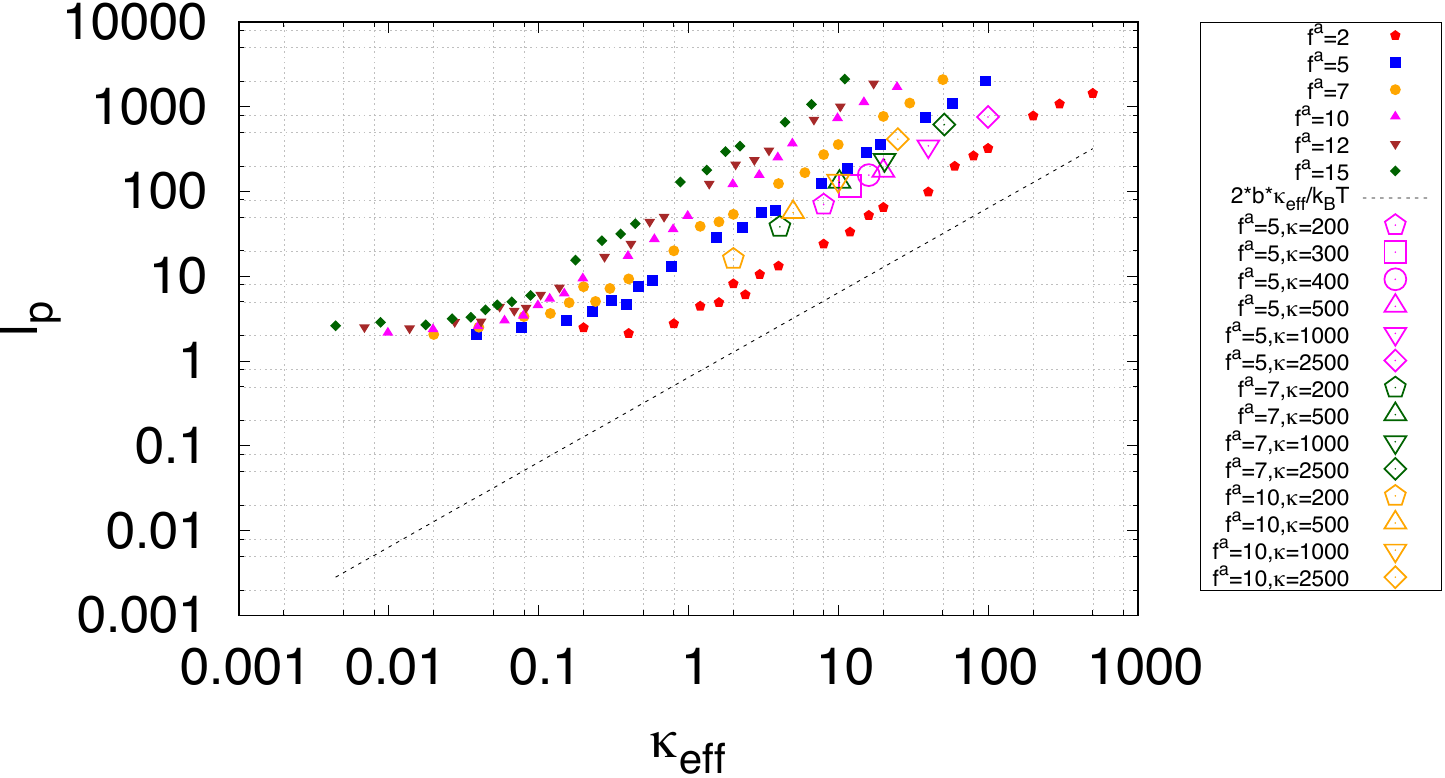}
\caption{\label{fig:kvslp_all} \rev{Effective persistence length $\lp$ estimated from the spectrum of filament tangent fluctuations in systems of isolated filaments (small filled symbols) and bulk active nematics (large open symbols). The dashed line has the slope of the expected scaling $2 b \ktp /\kt$. We see that all results scale with the bare bending rigidity,  $\lp \thicksim \kappa$, but only the results for the bulk systems exhibit data collapse for different activity strengths. The legend shows the value of $\kappa$ and $\fa$ for each system, and other parameters are $\tau_1=\tau$ and $k_\text{b}=30\ktRef/\sigma^2$.}
}
\end{figure}

At equilibrium, using Eqs. ~\eqref{eq:hbend_cont} and \eqref{eq:thetaq} along with the equipartition theorem results in
%
\begin{equation}
 \langle a_q^2 + b_q^2\rangle = \frac{2\kt}{\tilde{\kappa}L q^2}.
 \label{eq:ktilde}
\end{equation}
The modulus $\tilde{\kappa}$ can then be estimated from the slope of $\langle a_q^2 + b_q^2\rangle$ vs. $q$, as shown for an example parameter set in Fig.~\ref{fig:kvslp_all}b, and the persistence length is given by $l_\text{p} = \tilde{\kappa}/\kt$.

Performing this procedure for our non-equilibrium system as a function of $\kappa$ and $\fa$ allows estimating the activity-renormalized filament persistence length. Fig.~\ref{fig:kvslp_all}a
shows estimated persistence lengths as a function of $\kappa$ and $\fa$ measured from individual chains from two sets of simulations: isolated chains (small solid symbols) and chains from the bulk active nematic (large open symbols).  For chains within bulk, we observe data collapse when the effective persistence length is plotted as a function of $\ktp$ (provided $\fa\ge5$), with the same scaling found for the bulk bend modulus in the main text:  $l_\text{p} \thicksim \ktp$, consistent with the equilibrium result that the bulk bend modulus depends linearly on the filament modulus. In contrast, while the estimated persistence length scales linearly with $\kappa$ for isolated chains, we do not observe data collapse for isolated chains simulated at different activity strengths. This result supports the proposal in the main text that the observed scaling form for the activity-renormalized bend modulus arises due to inter-chain collisions. Furthermore, for $\fa\ge5$ the persistence length measured for isolated chains exceeds that of bulk chains, showing that the inter-chain collisions lead to an apparent softening of the filament modulus. 

\begin{figure}[h]
\vspace{1cm}
\includegraphics[width=0.8\linewidth]{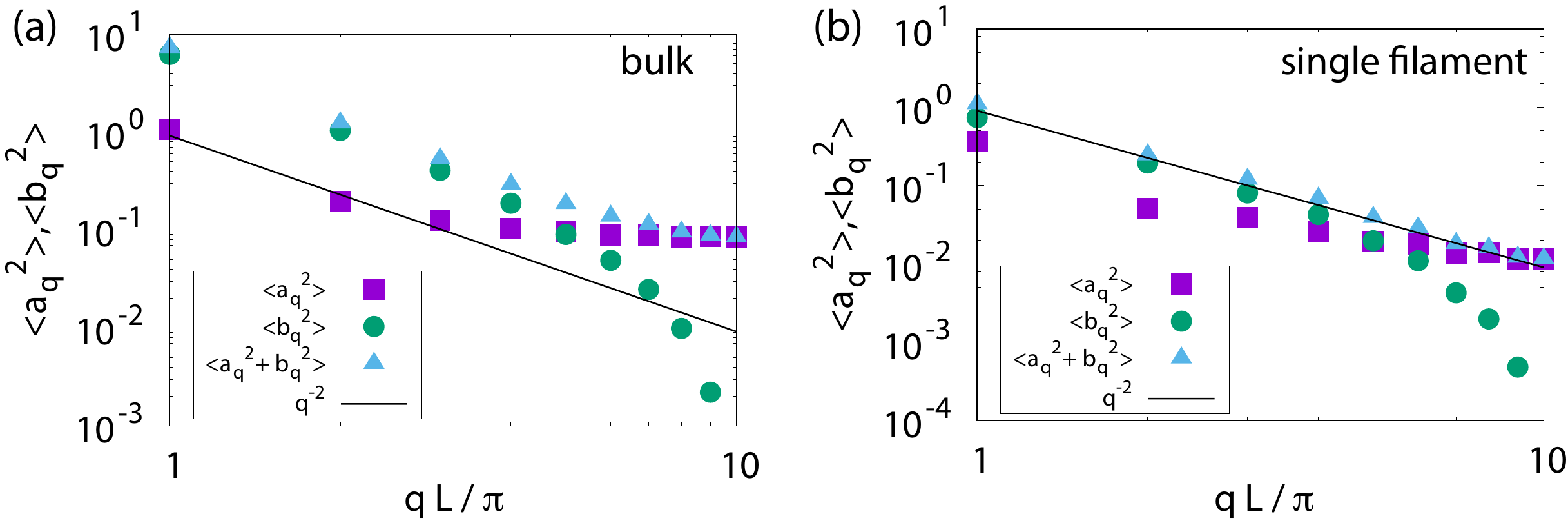}
\caption{\label{fig:abq} \rev{Fourier amplitudes $\langle a_q^2 \rangle$, $\langle b_q^2 \rangle$ and $\langle a_q^2 + b_q^2 \rangle$ as a function of wave vector $q L/\pi$, with $L$ the box side length, measured in a simulation of {\bf (a)} a bulk simulation and {\bf (b)} an isolated filament, for representative parameter values ($\kappa=500$,$\fa=10$;$\ktp=5$).}}
\end{figure}

\color{black}

\section{Splay and bend deformations}

\begin{figure}[h]
\includegraphics[width=0.7\linewidth]{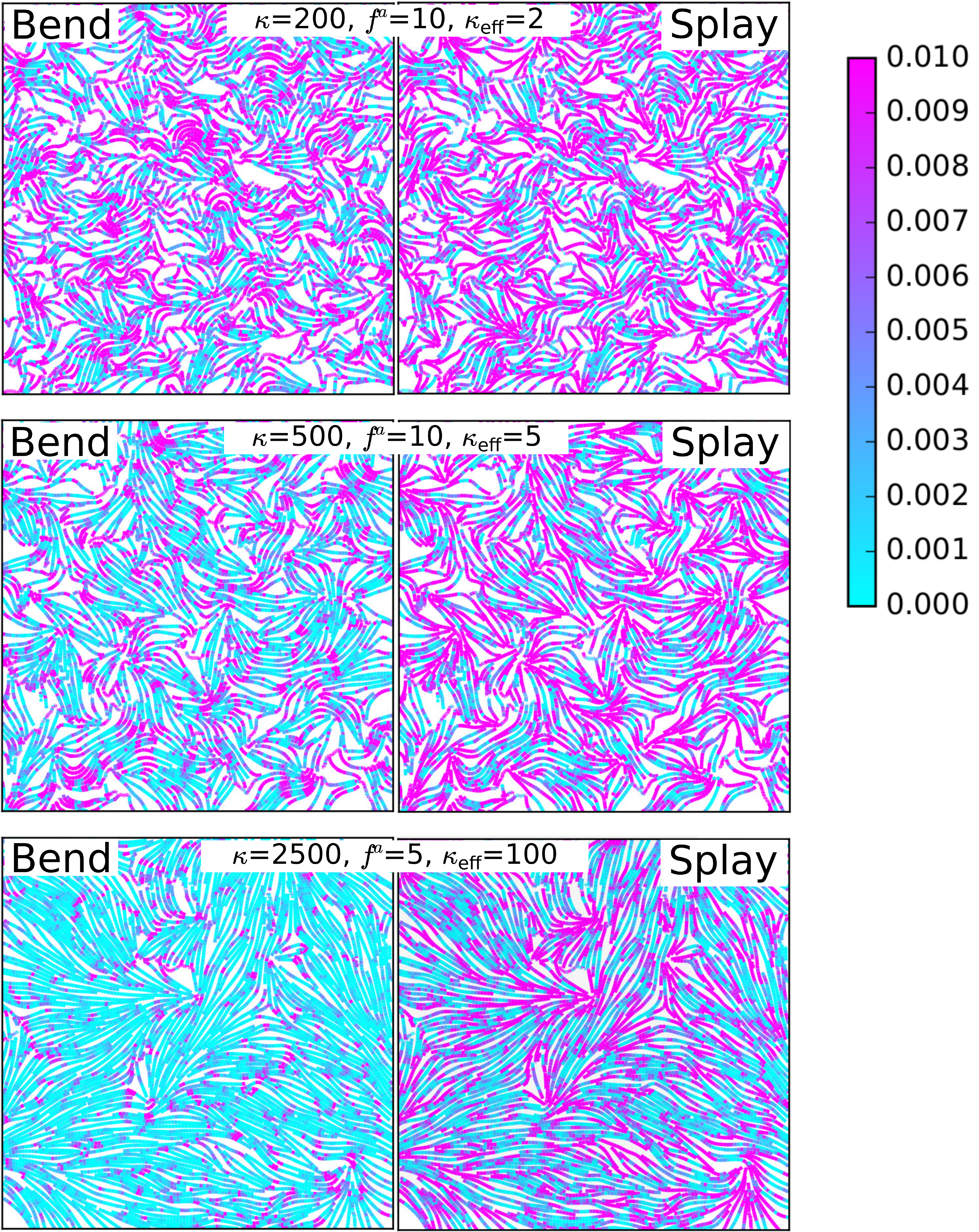}
\caption{\label{fig:bs_colormap} \rev{Snapshots from steady state configurations at indicated parameter values, with colormaps showing the distribution of bend, $\rho S^2(\hat{n}\times(\nabla\times\hat{n}))^2$ (left), and splay, $\rho S^2 (\nabla\cdot\hat{n})^2$ (right), superposed on lines representing the director field. The parameters are chosen to highlight differences between flexible (small $\ktp$)and
rigid (large $\ktp$) systems. The color range is clipped at 0.01 to to clarify the spatial variations of the deformations.}}
\end{figure}

\begin{figure*}[hbt]
\includegraphics[width=0.8\linewidth]{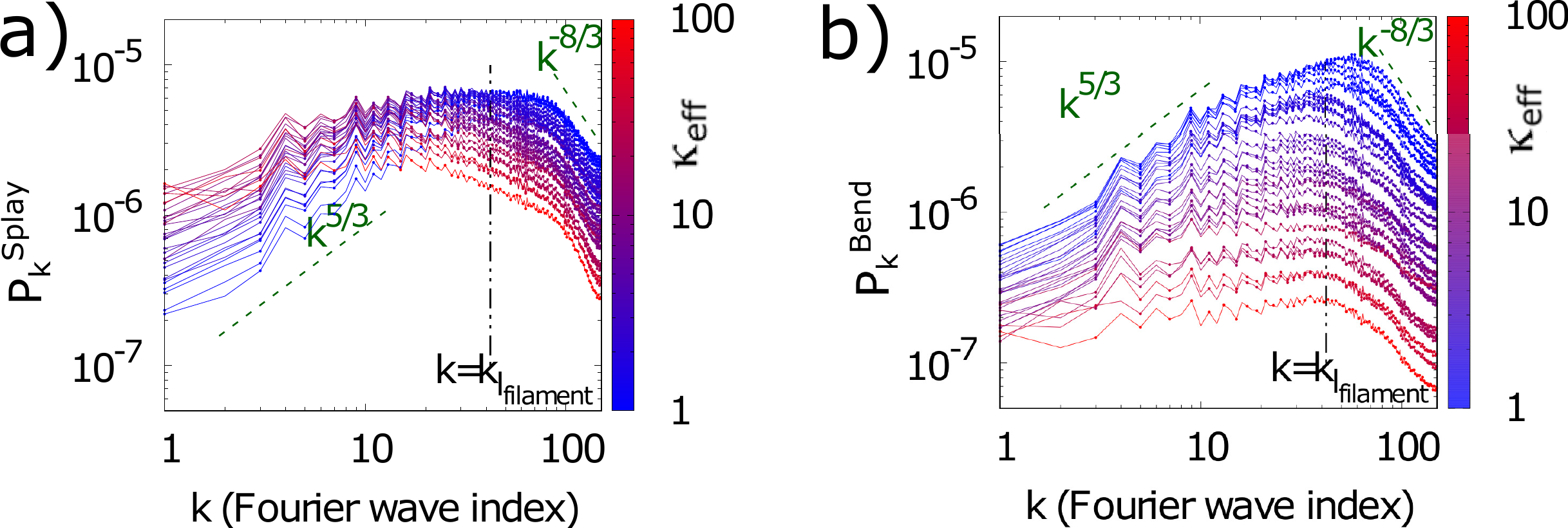}
\caption{\label{fig:power_spectrum} \rev{ Power spectrum  of splay {\bf (a)} and bend {\bf (b)} deformations as a function of wavenumber $k$,
with the colorbar indicating the value of effective bending rigidity ($\ktp$). }}
\end{figure*}


In a continuum description of a 2D nematic, all elastic deformations can be
decomposed into bend and splay modes,
given by $\dbend(\mr)=\left(\hat{n}(\mr) \times (\nabla \times \hat{n}(\mr)\right)$
and $\dsplay(\mr)=\left(\nabla\cdot\hat{n}(\mr)\right)$.
Fig.~\ref{fig:bs_colormap} shows the spatial distribution of bend and splay deformations in
systems at low and high rigidity values. To avoid breakdown of these definitions within defect
cores or other vacant regions, we have normalized the deformations by the local density
and nematic order: $\Dbend=\rho S^2 |\dbend|^2$ and $\Dsplay=\rho S^2 \dsplay^2$.
We see that bend and splay are equally spread out in the system in the
limit of low rigidity, whereas bend deformations are primarily located near defect
cores for large rigidity.  In the
high rigidity simulations, the effective persistence length ($\approx 66\sigma$) significantly exceeds the filament contour length ($20\sigma$),
and thus most bend deformations correspond to rotation of the director field around filament ends at a defect tip.

To obtain further insight into the spatial organization of deformations, we
 calculated power spectra as $P^\text{bend}_k = \int d^2 \mathbf{r}' \exp\left(-i \mathbf{k} \cdot (\mathbf{r}') \right) \langle \Dbend(\mr) \Dbend(\mr+\mathbf{r}')\rangle$, with
 an analogous definition for splay, and with the $Q$ field
 calculated at $1000\times1000$ grid points (a realspace gridspacing of $0.84\sigma$).
The resulting power spectra are shown in Fig.~\ref{fig:power_spectrum}a,b as functions of the renormalized filament rigidity, and the dependences of the peak positions and maximal power are discussed in the main text. Here we note that the splay spectra exhibit asymptotic scaling of $k^{5/3}$ and $k^{-8/3}$ at scales respectively
above the defect spacing or below the size of individual filaments, with a plateau region at intermediate scales.
The same assymptotic scalings in power spectra were observed in dense bacterial suspensions in the turbulent
regime~\cite{Wensink2012,Chatterjee2016}.

\begin{figure*}
\includegraphics[width=\linewidth]{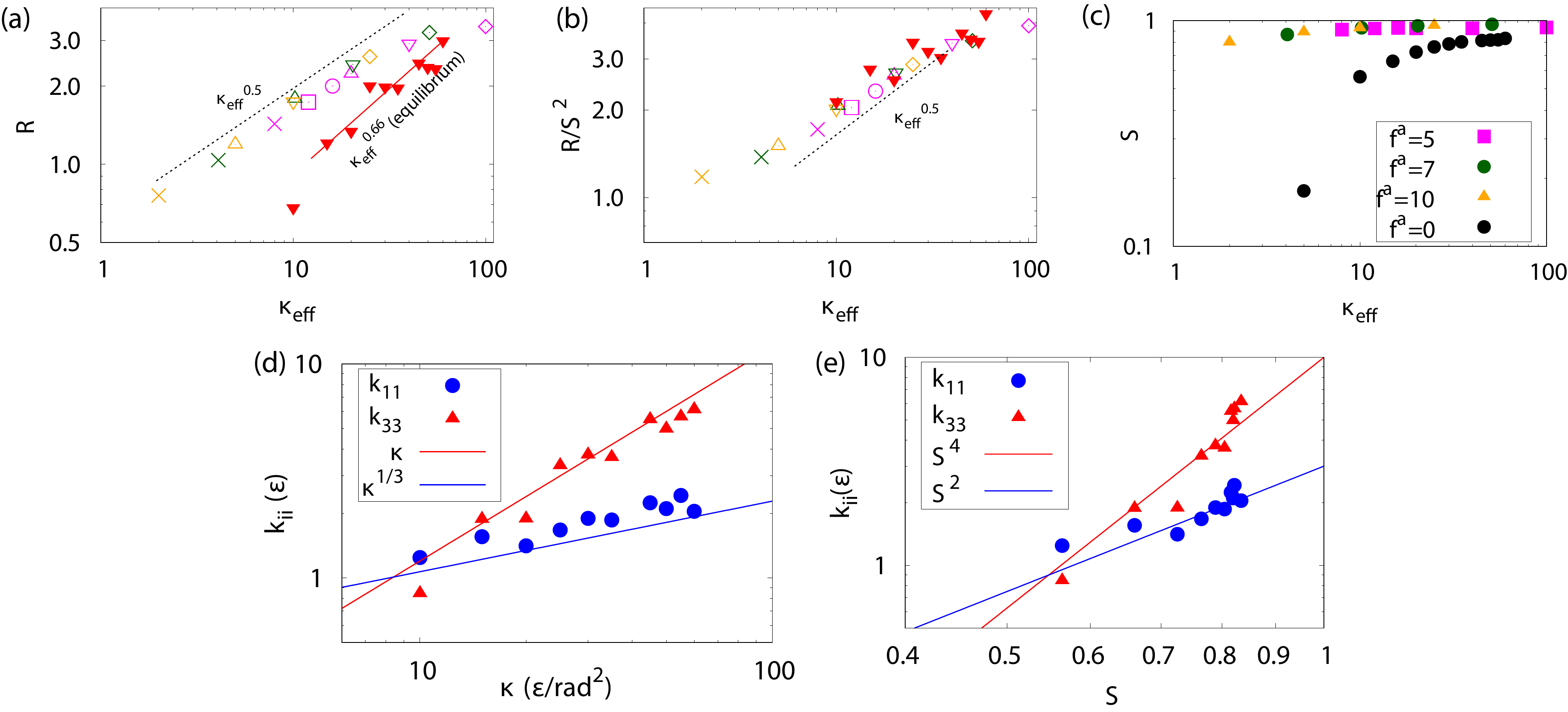}
\caption{\label{fig:kii} Comparison of ratio of bend and splay deformations in an active nematic to an equilibrium system.
{\bf (a)} The ratio of splay and
bend deformations, $R$,  plotted as
a function of renormalized bending rigidity for active and equilibrium systems. The ratio $R$ calculated in equilibrium systems is shown as \textcolor{red}{$\blacktriangledown$} symbols, while the symbols for the active system are defined as in Fig.~\ref{fig:old_fig_3}. {\bf (b)} The data from (a) is shown normalized by the mean nematic order parameter squared, (${\emph{R}}/S^2$).
{\bf (c)} The mean nematic
order parameter, $S$, as a function of renormalized bending rigidity measured in active and equilibrium simulations.
{\bf (d)} Values of the bend ($k_{33}$) and splay ($k_{11}$) elastic constants as a
function of $\kappa$ calculated using free energy perturbation \cite{Joshi2014} in equilibrium simulations ($\fa=0$).
{\bf (e)} The same results as in (d), plotted against the nematic order
parameter $S$ calculated for each parameter value. The red and blue lines
indicate scaling of $\sim S^4$ and  $\sim S^2$.
 \rev{The active results in this figure correspond to the additional data set with  $\kfene=30$ and $\tau_1=1$, as in Fig.~\ref{fig:old_fig_3}.}}
\end{figure*}

The main text discusses the ratio of total strain energy in splay
deformations to those in bend
\begin{align}
 {\emph{R}}=\left\langle \int d^2\mr \Dsplay(\mr)\right\rangle / \left \langle \int d^2\mr \Dbend(\mr)\right\rangle
 \label{Rsupp}.
 \end{align}
We find that this ratio scales as ${\emph{R}}\thicksim \ktp^{1/2}$ for all
parameter sets, which is different from the expected scaling in an
equilibrium nematic of ${\emph{R}}_\text{eq}\thicksim \kappa^{2/3}$. To
investigate the origins of this discrepency, we measured the elastic moduli
for an equilibrium system for different $\kappa$ values shown in Fig.~(\ref{fig:kii}).
We find that the degree of order in the
system depends on the value of $\kappa$,  that approximate scalings can be
identified as $k_{33}\sim S^4$ and  $k_{11}\sim S^2$ (Fig.~(\ref{fig:kii}b)), and that the amount of order
in the system at a given stiffness value $\ktp$ is very
different for active nematics when compared to their equilibrium analogs (Fig.~(\ref{fig:kii}c)).
Active nematics have considerably higher order,
possibly because their intrinsic tendency to phase
separate \cite{Mishra2006,Chate2006} leads to higher local density in comparison
to an equilibrium system at corresponding $\rho$ and $\ktp$. We used this information to empirically find that
 ${\emph{R}}/S^2$
exhibits approximately the same scaling for active and passive nematics.

\begin{figure}[hbt]
\includegraphics[width=0.5\linewidth]{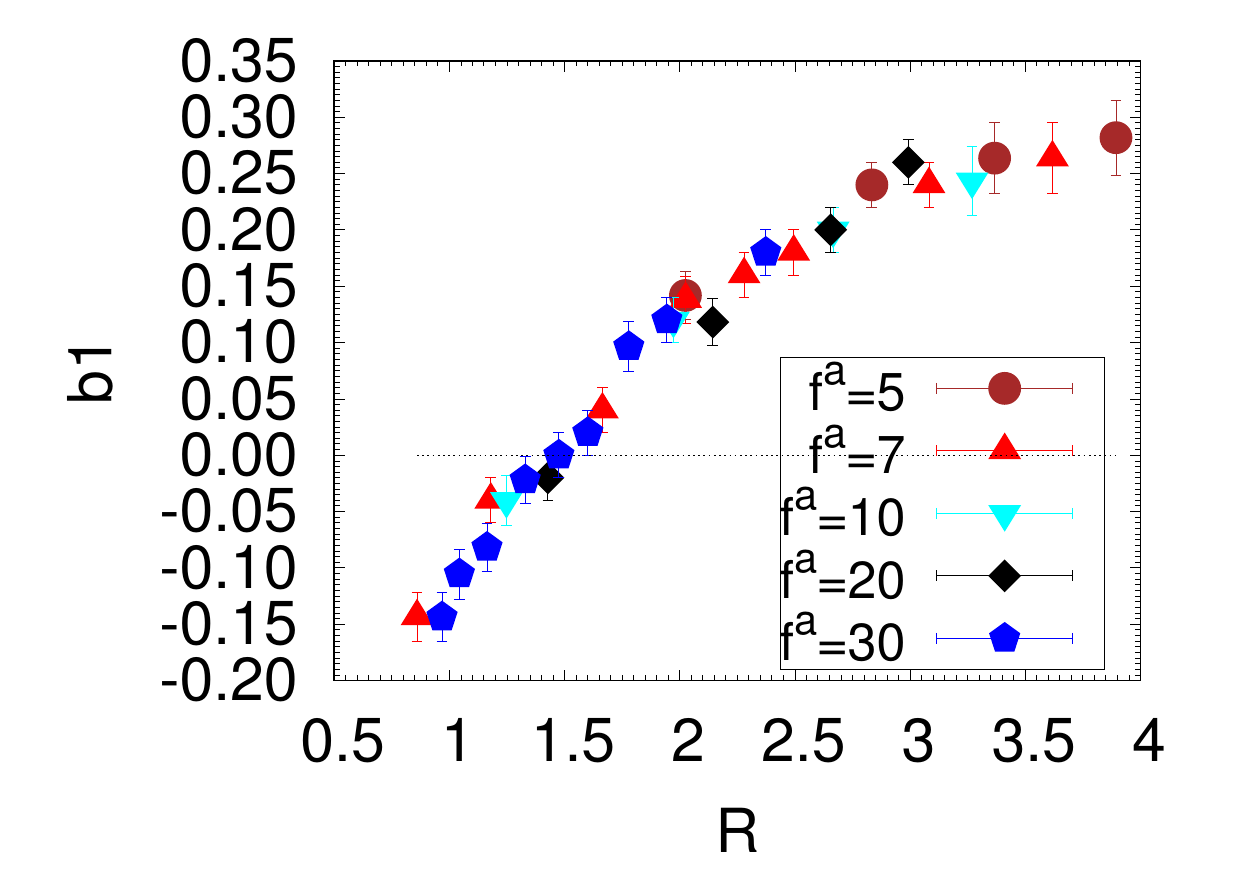}
\caption{\label{fig:sb_b1} \rev{Defect shape parameter $b_1$ as a function of splay/bend ratio $R$ defined in the text, for the data in Fig. 3b,c of the main text. }
}
\end{figure}

\rev{Finally, by analogy to equipartition at equilibrium, the ratio of splay/bend, $R$, can be construed as an effective ratio of moduli: $k_{33}^\text{effective}/k_{11}^\text{effective}$, with the ratio depending  on activity. Fig.~\ref{fig:sb_b1} shows the defect shape parameter plotted as a function of this ratio.}

\begin{figure}[h]
\vspace{1cm}
\includegraphics[width=\linewidth]{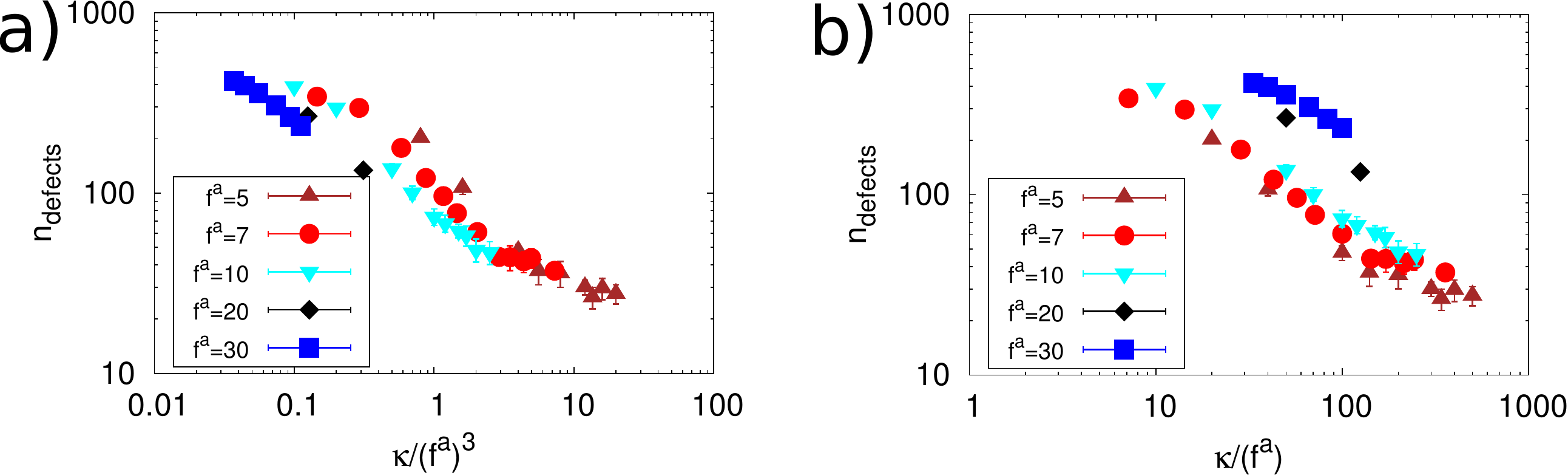}
\caption{\label{fig:dd_scales} \rev{The data from Fig. 3 (main text) for defect densities measured in simulations at varying $\kappa$ and  $\fa$ is plotted according to two alternative scaling forms for the activity-renormalized bending rigidity: {\bf (a)} $\ktp=\kappa/(\fa)^3$ and {\bf (b)} $\ktp=\kappa/\fa$. 
}
}
\end{figure}

\rev{\section{Testing the scaling form for the effective bending rigidity}
Fig. ~\ref{fig:dd_scales} shows two alternative scaling forms for the activity-renormalized bending rigidity $\ktp$, with data from a wide range of activity values ($(\fa)^2\in[25,900]$). We see that only the form presented in the main text, $\ktp\cong\kappa/\fa^2$ (Fig.~3 main text) leads to data collapse from simulations with different activity levels. }

\color{black}
\section{System size effects}
\begin{figure}[h]
\includegraphics[width=0.5\linewidth]{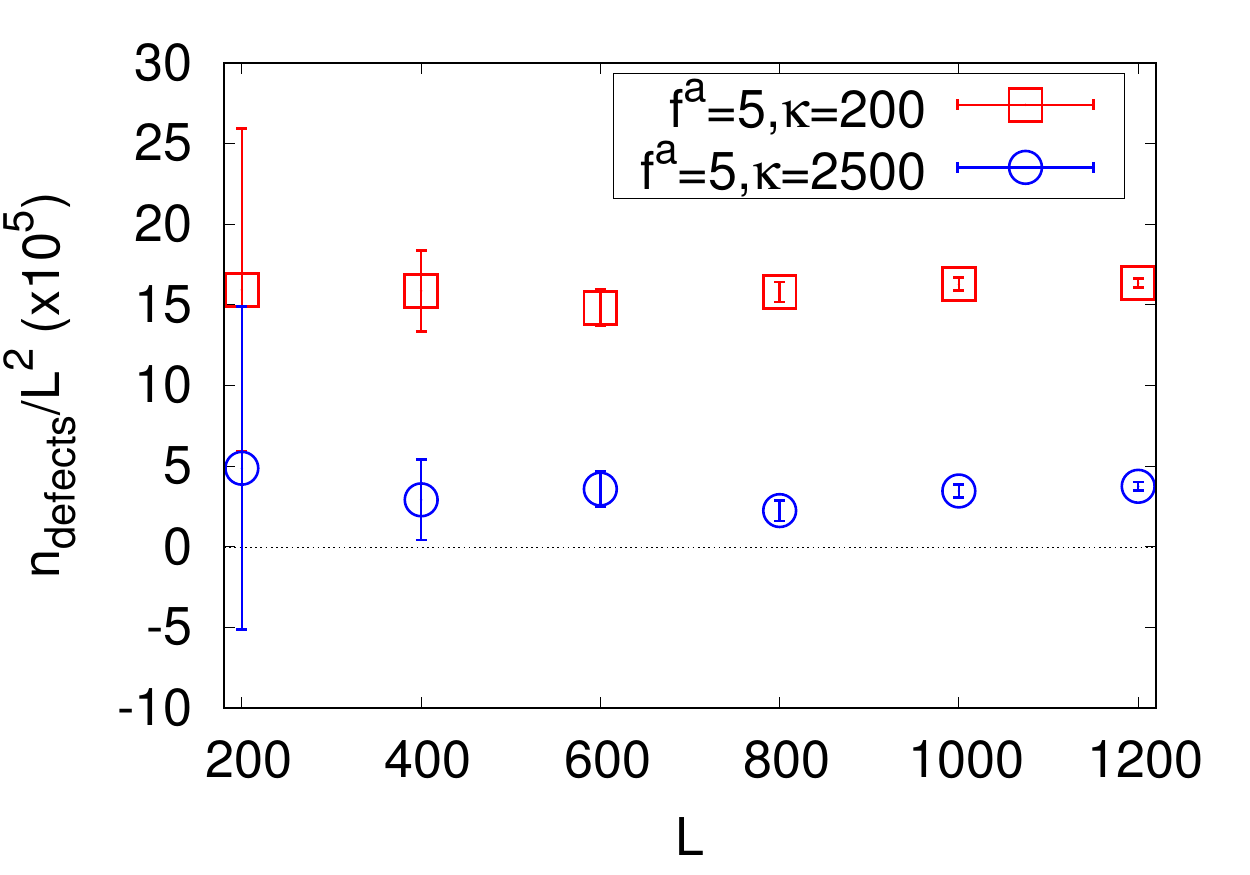}
\caption{\label{fig:system_size} \rev{Steady-state defect density as a function of simulation box side length $L$ for square boxes with periodic boundary conditions, at indicated values of the filament modulus and activity parameter, with $\tau_1=\tau$ and $k_\text{b}=30\ktRef$. }
}
\end{figure}

To assess finite size effects on our results, we performed a system size analysis for two parameter sets from Fig.~\ref{fig:old_fig_3}:  $\kappa=200,\fa = 5$ and $\kappa =2500,\fa=5$, with $\tau_1=\tau$ and $k_\text{b}=30$. We chose these parameter sets because they are near the  upper and lower limits of effective bending rigidity investigated in that set of simulations. As shown in Fig.~\ref{fig:system_size}, we observe no systematic dependence of defect density on system size over the range of side lengths $L\in[200,1200]\sigma$. We observe a similar lack of dependence on system size for other observables, suggesting that finite size effects are negligible in our simulations at system size ($840\times840 \sigma^2$).
\color{black}

\begin{figure}[hbt]
\includegraphics[width=0.8\linewidth]{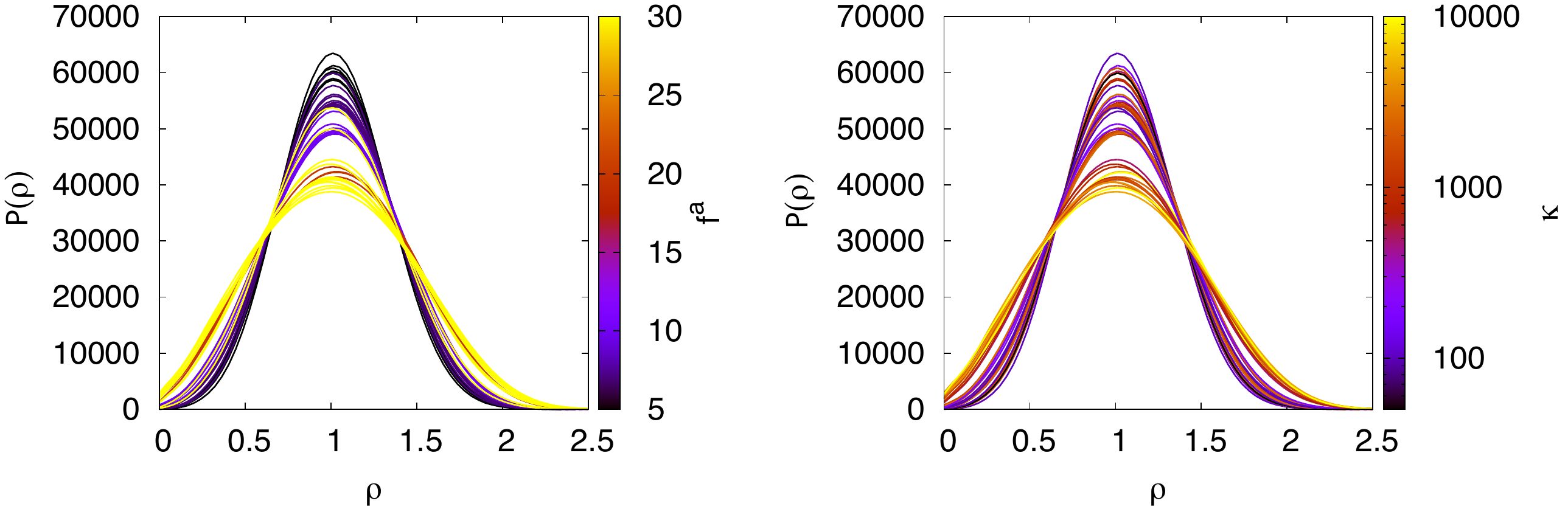}
\caption{\label{fig:rhostat_fa_kappa} \rev{The distribution of local densities of filament pseudoatoms for indicated values of activity \textbf{(left)} and bare bending rigidity \textbf{(right)}. Local densities were calculated by measuring the number of filament beads within square subsystems with side length $10\sigma$.}
}
\end{figure}

\rev{\section{Density fluctuations}}
\rev{It is well-known that active nematics are susceptible to phase separation \cite{Mishra2006,Ngo2014,Putzig2014,Shi2013} and giant number fluctuations (GNFs) \cite{Ramaswamy2003,Mishra2006,Chate2006,Narayan2007,Zhang2010}. We therefore monitored these quantities in our system. Interestingly, while we do observe large density fluctuations on small scales (see videos of typical trajectories), phase separation is suppressed on large scales in the semiflexible regime. Fig.~\ref{fig:rhostat_fa_kappa} shows histograms of local density, measured within subsystems with side length $10 \sigma$ as a function of $\kappa$. We see that the distribution of local densities broadens as $\fa$ and $\kappa$ increase, but remains unimodal indicating an absence of true phase separation. }

\begin{figure}[hbt]
\includegraphics[width=0.99\linewidth]{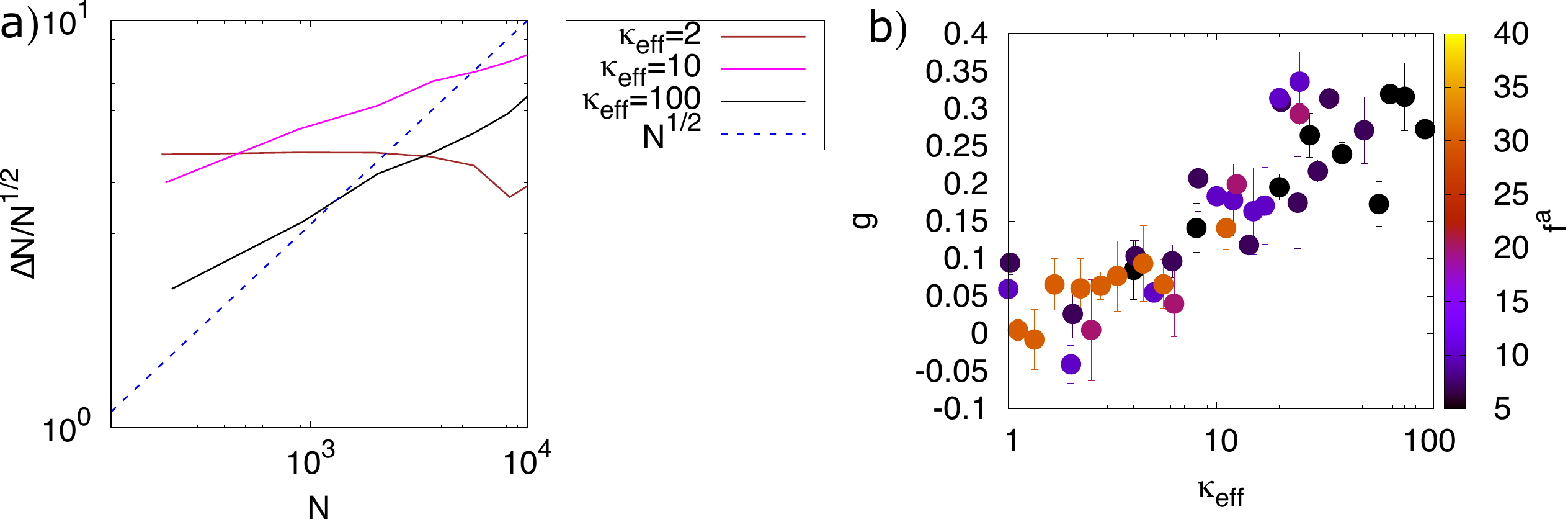}
\caption{\label{fig:giant} \rev{ Giant number fluctuations (GNFs) depend on the renormalized bending rigidity. \textbf{(a)}. The mean fluctuations of number of filament pseudoatoms, $\Delta N$, is plotted as a function of subsystem size $N$ for representative values of $\ktp$. The fluctuations are normalized by their value at equilibrium, $\Delta N / \sqrt{N}$ so that a horizontal line indicates equilibrium-like fluctuations. \textbf{(b)} The fluctuation scaling exponent $g$ is plotted as a function of $\ktp$. We determined $g$ by fitting the dependence of fluctuations on subsystem size for each parameter set to the form $\Delta N/\sqrt{N}=a N^g$ over the range $N\le 10^4$. Note that a value of $g=0$ corresponds to equilibrium-like fluctuations, while $g=0.5$ would correspond to $\Delta N \thicksim N$.
}
}
\end{figure}

\rev{We also monitored the existence of GNFs, by measuring the number of pseudoatoms within square subsystems with side lengths ranging from $20 \sigma$ to $840 \sigma$ as a function of time. At equilibrium, in a region containing on average $N$ particles the standard deviation of the number of particles $\Delta N$ scales as $\Delta N\thicksim \sqrt{N}$, while previous studies of active nematics have identified higher scaling, as large as $\Delta N\thicksim N$. Consistent with the other characteristics of an active nematic studied in this article, we find that the scaling of number fluctuations depends only on the effective bending modulus $\ktp$. Fig.~\ref{fig:giant}a shows measured number fluctuations for different values of the effective bending modulus,  plotted as $\Delta N / \sqrt{N}$, so that the result will be constant with subsystem size for a system exhibiting equilibrium-like fluctuations. We see that for small $\ktp=2$ the result is constant with subsystem size, indicating equilibrium-like fluctuations, but the slope increases for $\ktp=10$ and $\ktp=100$ indicating a progression toward GNFs. At all $\ktp$ the fluctuations are eventually suppressed on scales  comparable to the defect spacing ($N \gtrsim 10^4$), at which scale the system is essentially isotropic, consistent with Narayan \etal \cite{Narayan2007}.}

\rev{To determine the dependence on $\ktp$, we fit the data for each simulation in the range $N\le10^4$ to the form $a N^g$, so that $g=0$ indicates equilibrium-like fluctuations and $g=0.5$ would indicate linear scaling of fluctuations with system size. As shown in Fig.~\ref{fig:giant}b,  $g$ increases with $\ktp$, with $g=0$ for small $\ktp$ and $g \approx 0.3$ for the largest renormalized bending rigidity values investigated; \ie $\Delta N \thicksim N^{0.8}$. The fact that $g<0.5$ for the parameters we consider may reflect suppression of fluctuations even for $N<10^4$. Importantly, estimated values of $g$ at different  $\fa$ and $\kappa$ collapse onto a single function of $\ktp$, consistent with the observations of other characteristics (Fig. 3 in the main text).  We speculate that  we do not observe GNFs for small renormalized bending rigidity values because GNFs are suppressed on scales beyond the defect spacing.
}

\section{Defect identification and shape measurement algorithm }
\label{measuring_defect_shape}
\begin{figure}[h]
\vspace{1cm}
\includegraphics[width=0.3\linewidth]{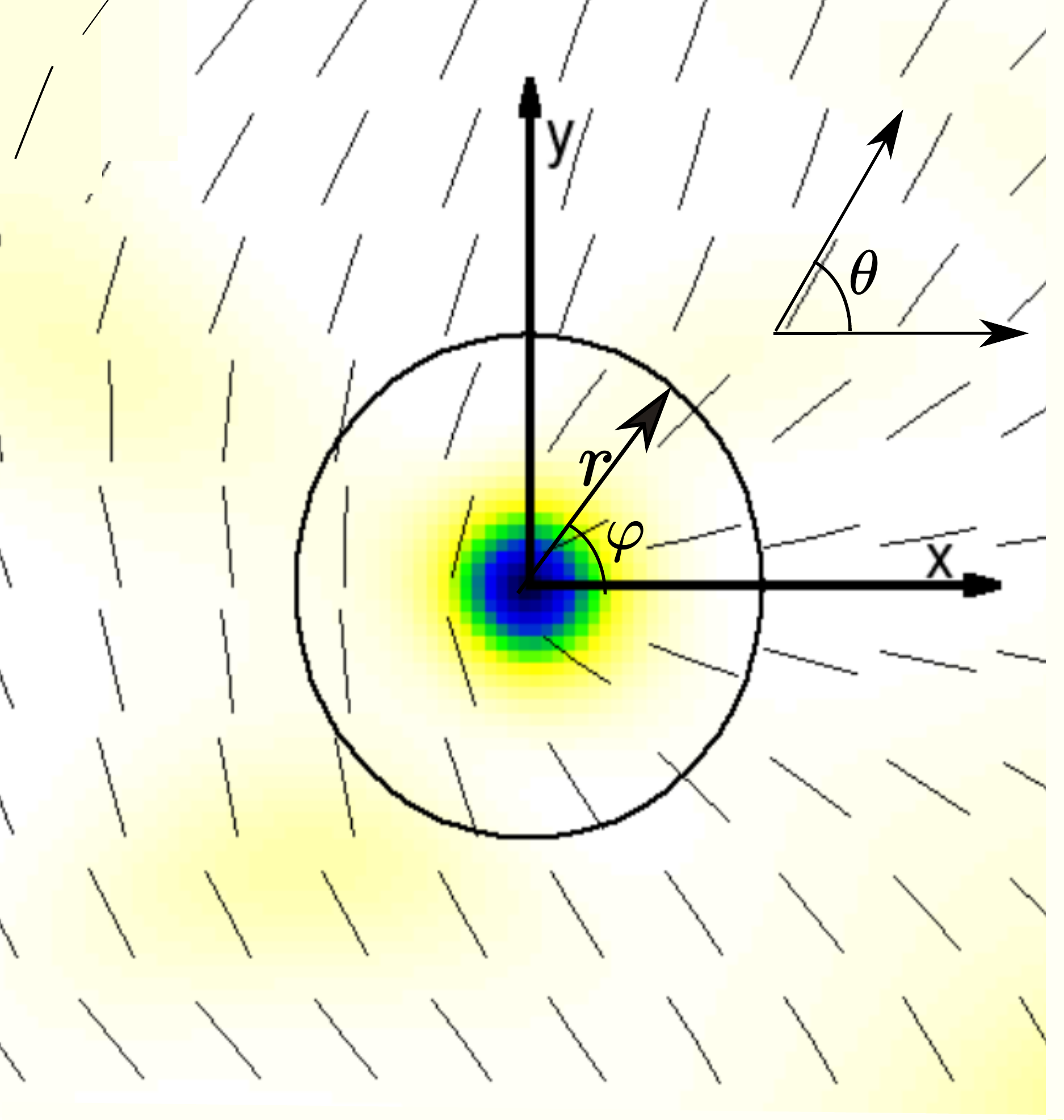}
\caption{\label{fig:plus_half_illust} Schematic showing the defect-centered coordinate system defined in the text, with azimuthal angle $\phi$ and director angle $\theta(r,\phi)$. Note that the $x$-axis is chosen along the orientation vector of the +1/2 defect given by the angle $\theta_{0}'$ defined in the text.}
\end{figure}

Here, we provide details on how we identify and measure the shapes of defects from our simulation data. This algorithm can also be directly applied to retardance images from experimental systems, and discretized output from continuum simulations.

\underline{Locating and identifying defects}: We locate defects using the fact the magnitude of nematic order $S$ is very small at defect cores. We first compute the magnitude $S=2\sqrt{Q_{xx}^{2}+Q_{xy}^{2}}$ from the nematic tensor, whose measurement was described above. The regions corresponding to defect cores can then be extracted by using a flood-fill algorithm to select connected areas where the order is below some
threshold $S_\text{threshold}$. We set $S_\text{threshold}=0.6$ since the system is deep within the nematic state for the parameters of this study.
Once the defect cores have been located, the charge of each defect can be identified
from the total change in the orientation of the director in a loop around
the defect core. We perform this calculation by adding the change in angle for points
in a circle about the center of the core. We choose the radius of the circle to be at least
$5\sigma$, to ensure a well defined director field. The total change in angle must be a multiple of $\pi$:
$\Delta\theta=n\pi$, where if $n=0$ then the disordered region is
not a defect, and otherwise it is a defect with topological charge $m=\frac{n}{2}$.
Typically $n=\pm1$ but, in rare cases we observed defects with charge $m=+1$ in our simulation data.

\underline{Identifying the orientations and characterizing the shapes of $\phd$ defects}: Given the location of a defect and its charge, there are several
methods which can be used to measure the orientation of the $\phd$ defects \cite{DeCamp2015,Vromans2015}. In this work, we compute the  sum of the divergence of ${\bf Q}$ field, $\nabla_{\beta}  Q_{\alpha \beta}$ along a circle enclosing the defect, and normalize it to a give unit vector. This unit vector represents the orientation of the +1/2 defect and in our two dimensional system identifies an angle $\theta'_{0}$ for the defect.

We then measure the director orientation $\bar{\theta}(\bar{\phi})$ along the azimuthal angle $\bar{\phi}$
at discrete set of radii, $\{r\}$, around the defect core. First, we ensure that each loop does not cross any disordered regions, or enclose any other defects, by checking
the order at each point and summing $\Delta\theta$ over the loop.
Then we apply a coordinate frame rotation such that $\theta=\bar{\theta}-\theta'_{0}$, and the azimuthal angle $\phi=\bar{\phi}-\theta'_{0}$, where, $\theta'_0$ is an orientation of the +1/2 defect estimated above. This step rotates the coordinate frame of reference to the frame of reference of the +1/2 defect. Finally, we evaluate the Fourier coefficients
for $\theta(\phi)$,
\begin{equation}
\theta(r,\phi)=\phd \phi + \sum_n a_n(r) \cos(n\phi) + b_n(r) \sin(n\phi).\label{eq:theta_r_phi}
\end{equation}
However, in practice we find that truncating the
expansion after the first sin term gives an excellent approximation of the
shape of a $\phd$ defect. \rev{Hence, once a value of $r$ is chosen,} the defect can be characterized by the single parameter $b_1$.

\begin{figure}[hbt]
\includegraphics[width=0.45\linewidth]{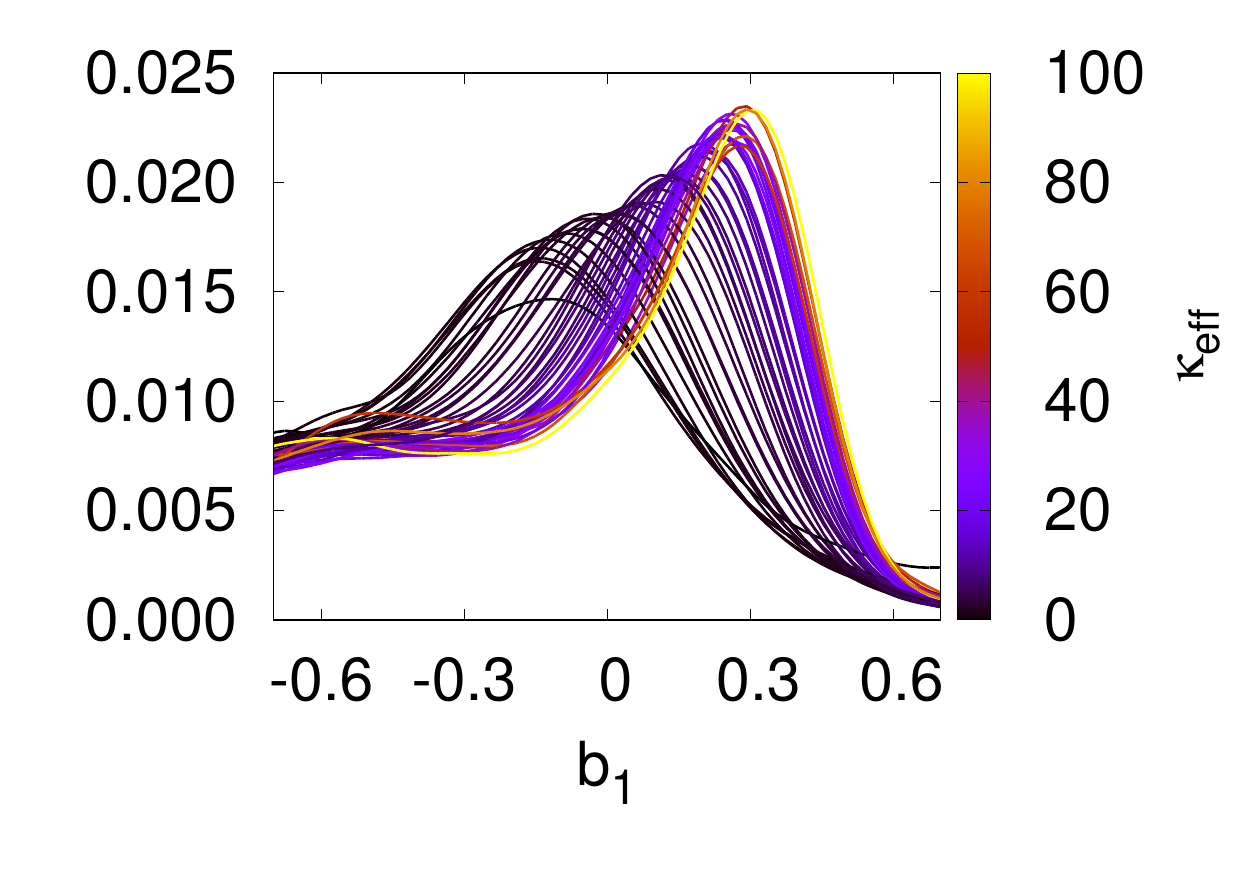}
\includegraphics[width=0.45\linewidth]{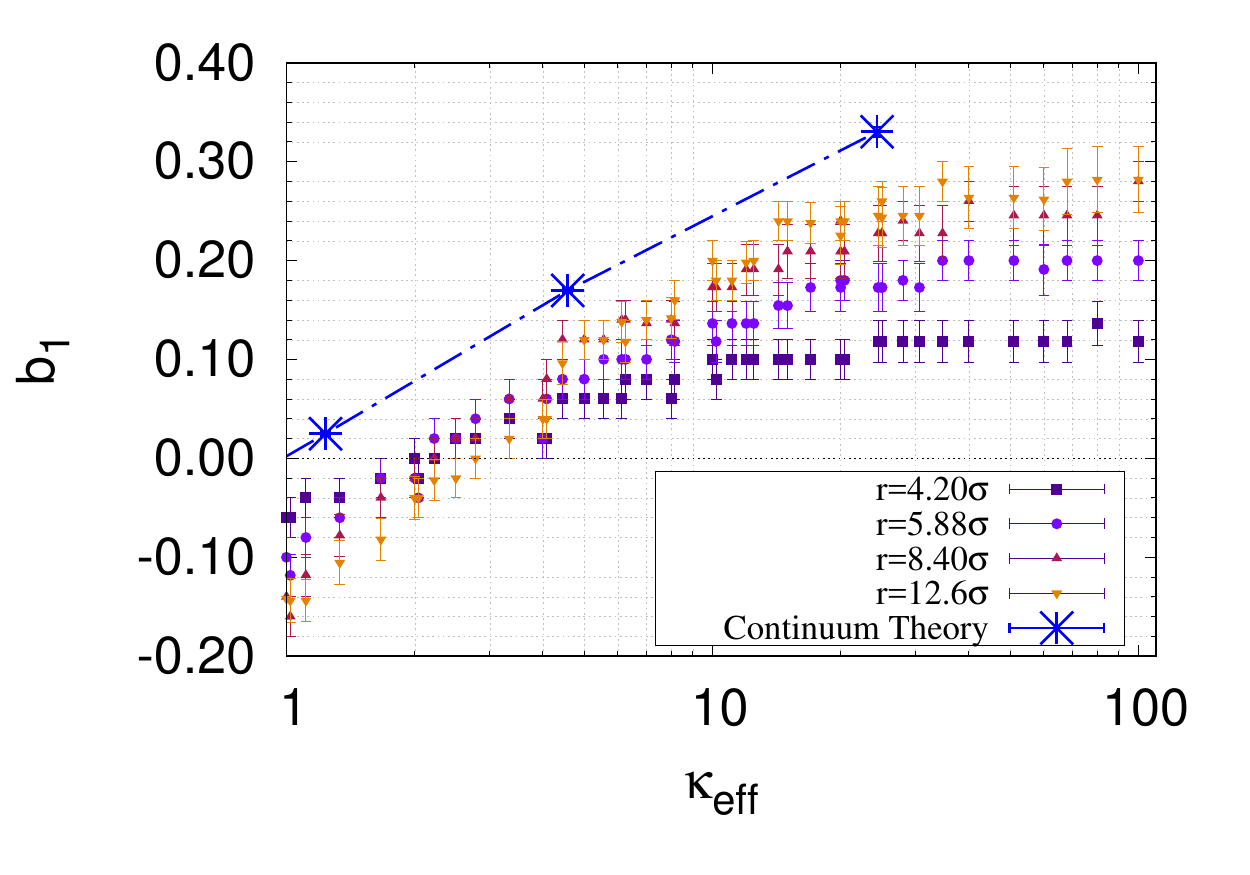}
\caption{\label{fig:b1hist_new} \rev{\textbf{(left)} Normalized distributions of $b_1$ obtained from the simulations in Fig. 3 (main text), with parameterizations calculated at a distance $r=12.6\sigma$ from the center of each defect according to Eq.~\eqref{eq:theta_r_phi}. \textbf{(right)} Effect of varying $r$ (the distance from the center of the +1/2 defect core) on the defect shape parameter $b_1$. The mode of $b_1$ is shown as a function of $\ktp$ for the simulations  in Fig.~3 (main text), with indicated values of $r$.}
}
\end{figure}


In Fig.~\ref{fig:b1hist_new} we show the distribution of $b_1$ values obtained from our simulations {with $r=12.6\sigma$}. Note that we observe long tailed distributions of $b_1$ with tails in the $b_1 < 0 $ regime. However, the distributions are sharply peaked with typical peak width $\sim 0.1$. Therefore we consider the mode of $b_1$ values as an appropriate measure of defect shape.

\rev{\underline{Choice of $r$}:
For an isolated defect, $b_1$ asymptotes once the distance from the defect center increases beyond the core size. However, as noted in Zhou et al. \cite{Zhou2017}, in a system with finite defect density the defect shape should be parameterized as close to the defect center as possible to avoid distortion due to other defects. The typical defect core radius in our simulations (defined as the region in which the nematic order parameter $S < 0.6$) is about $4\sigma$. The smallest defect spacing (at the highest defect density) in our simulations is about $40 \sigma$. We therefore chose a radius $r=12.6\sigma$, where the nematic is highly ordered and the director is always well-defined, but distortions due to other defects are minimized. As shown in Fig.~\ref{fig:b1hist_new}b the results are  qualitatively insensitive to radius for $r > 5$, although statistics become more limited for larger $r$. For consistency, the same radius should be chosen for all systems.}


\rev{\underline{Breakdown at high $\fa$ and $\kappa$}: As noted in the main text, the defect identification algorithm breaks down in systems with both extremely high activity and high bare bending rigidity ($\fa \ge 20$ and $\kappa\gtrsim 5000$). Under these conditions the system exhibits density fluctuations on very short length scales (see Fig.~\ref{fig:rhostat_fa_kappa} and the movie showing snapshots from a simulation trajectory with $\fa=30$ and $\kappa=10^4$). The defect algorithm cannot distinguish between configurations in which stiff rods trans-pierced these holes and actual defects. Therefore we have not measured defect densities for these parameter sets.}

\bibliography{dshapebib}